\documentclass[onecolumn,preprintnumbers,amsmath,amssymb]{revtex4}
\usepackage{graphicx}
\usepackage{amsfonts}
\usepackage{amssymb}
\usepackage{rotating}
\usepackage{booktabs}
\usepackage{xcolor}
\usepackage{soul}
\usepackage{color}
\usepackage{slashed}
\usepackage{multirow}
\usepackage{makecell}
\usepackage{epsf}
\usepackage{ulem}
\usepackage{cancel}
\usepackage{color,bm}
\usepackage[colorinlistoftodos]{todonotes}
\usepackage[colorlinks=true,citecolor=cyan,urlcolor=blue,bookmarks=true,bookmarks=true,bookmarksopen=true,bookmarksnumbered=true,bookmarksopenlevel=3]{hyperref}

\usepackage{diagbox}

\definecolor{airforceblue}{rgb}{0.36, 0.54, 0.66}
\definecolor{steelblue}{rgb}{0.27, 0.51, 0.71}
\definecolor{amber}{rgb}{1.0, 0.49, 0.0}

\def\comment#1{}

\begin{document}

\title{Hard diffractive $\rm \eta_{c,b}$ hadroproduction at the LHC}
\date{\today}
\author{\textsc{Tichouk}}
\author{\textsc{Hao Sun}\footnote{Corresponding author: haosun@mail.ustc.edu.cn \hspace{0.2cm} haosun@dlut.edu.cn}}
\author{\textsc{Xuan Luo}}
\affiliation{Institute of Theoretical Physics, School of Physics, Dalian University of Technology, \\ No.2 Linggong Road, Dalian, Liaoning, 116024, P.R.China }

\begin{abstract}

In this paper, we investigate the inclusive diffractive hadroproduction for  $\rm \eta_{c}$ and $\rm \eta_{b}$ at the LHC energies. Based on the NRQCD factorization formalism and the resolved-Pomeron model for the quarkonium production mechanism, we estimate the rapidity, momentum fraction loss dependence of the cross section. We give prediction ratios for single and central diffractive processes with respect to non diffractive process. These inclusive processes are sensitive to gluon content of Pomeron for small-$x$ and Reggeon for large-$x$, useful to study small and large-$x$ physics and good to test different mechanism for $\rm \eta_{c}$ and $\rm \eta_{b}$ production at the LHC. They also serve as the background to related exclusive processes thus should be predicted. Our results demonstrate that the Reggeon contribution of diffractive processes can be sizable, even sometimes dominant over Pomeron, and that its study can be useful to better constrain the Reggeon parton content. The experimental study of Reggeon can be carried out in certain kinematic windows.

\vspace{0.5cm}
\end{abstract}
\maketitle
\setcounter{footnote}{0}

\section{INTRODUCTION}
\label{Intro}

The quarkonia production remains a topic of considerable theoretical and experimental interests in hadronic collisions at the Large Hadron Collider (LHC) and has attracted a lot of attention. It provides a valuable tool to test the ideas and methods of the QCD physics of bound states, such as effective field theories, lattice QCD, NRQCD, and so on \cite {HarlandLang:2009qe}. Recently, the $\rm \eta_{c}$ hadroproduction cross section was measured by the LHCb experiments \cite{Aaij:2014bga,Aaij:2019gsn} in $pp$ collisions which opened a window for the study of the pseudoscalar quarkonia production. This released experimental data provides a further important test for theories predicting the $\rm \eta_{c}$ hadroproduction cross sections polarisation. The investigation of direct $\rm \eta_{c}$ hadroproduction at leading order (LO) in $\alpha_{s}$ within NRQCD framework in the collinear factorization scheme has been carried out in Refs\cite{Biswal:2010xk,Likhoded:2014fta,Mathews:1998nk,Hao:1999kq} to describe the heavy quarkonium measurements. Besides, the QCD next-to-leading order (NLO) predictions of direct $\rm \eta_{c}$ hadroproduction have achieved good agreement with almost all the experimental measurements on quarkonia hadroproduction and clarified the ambiguity of the determination of the color octet long distance matrix elements for $\rm J/\psi$ production \cite{Zhang:2014ybe,Feng:2019zmn,Lansberg:2017ozx}. However, the notorious $\rm J/\psi$ polarization in hadroproduction became rather puzzling for conventional nonrelativistic QCD (NRQCD) calculations at NLO in comparison to the world's data with transverse momenta up to 10 GeV \cite{Butenschoen:2014dra}. Moreover, the examination of small or no polarization in $\rm J/\psi$ meson prompt production \cite{Aaij:2013nlm} remains mysterious within the accessible theoretical framework\cite{Brambilla:2010cs}. The theory also lost its flexibility and made a prediction for $\rm \eta_{c}$ by a huge factor off the measured cross section with  nonperturbative matrix elements  fixed from fitting all other production data \cite{Han:2014jya}. The general situation was even known as challenging \cite{Butenschoen:2014dra}. The investigation of the ground state, the $\rm \eta_{c}$ meson is still required so as to offer useful additional information on the long-distance matrix elements \cite{Butenschoen:2012px,Chao:2012iv} and particularly, the heavy-quark spin-symmetry relation between the $\rm \eta_{c}$ and $\rm J/\psi$ matrix elements. Therefore, more studies on the $\rm \eta_c$ productions are being worked on or have been published, see, for example, the prompt $\rm \eta_{c}$ heavy quarkonium production that has been intensively examined in the TMD-factorization with transverse momentum dependent distributions of on-shell gluons\cite{Echevarria:2019ynx}, and the $\rm K_{T}$-factorization scheme \cite{Baranov:2019joi} along with the potential model in the transverse momentum space with off-shell gluons \cite{Babiarz:2019mag}.

It is not surprised that the $\rm \eta_c$ hadroproduction can be also used to study the soft interactions at the LHC, for example, through diffractive production modes in which no quantum numbers are exchanged between colliding particles at high energies. 
\begin{figure}[htbp]
\centering
%\begin{minipage}[t]{4.0cm}
\includegraphics[height=3.4cm,width=3.4cm]{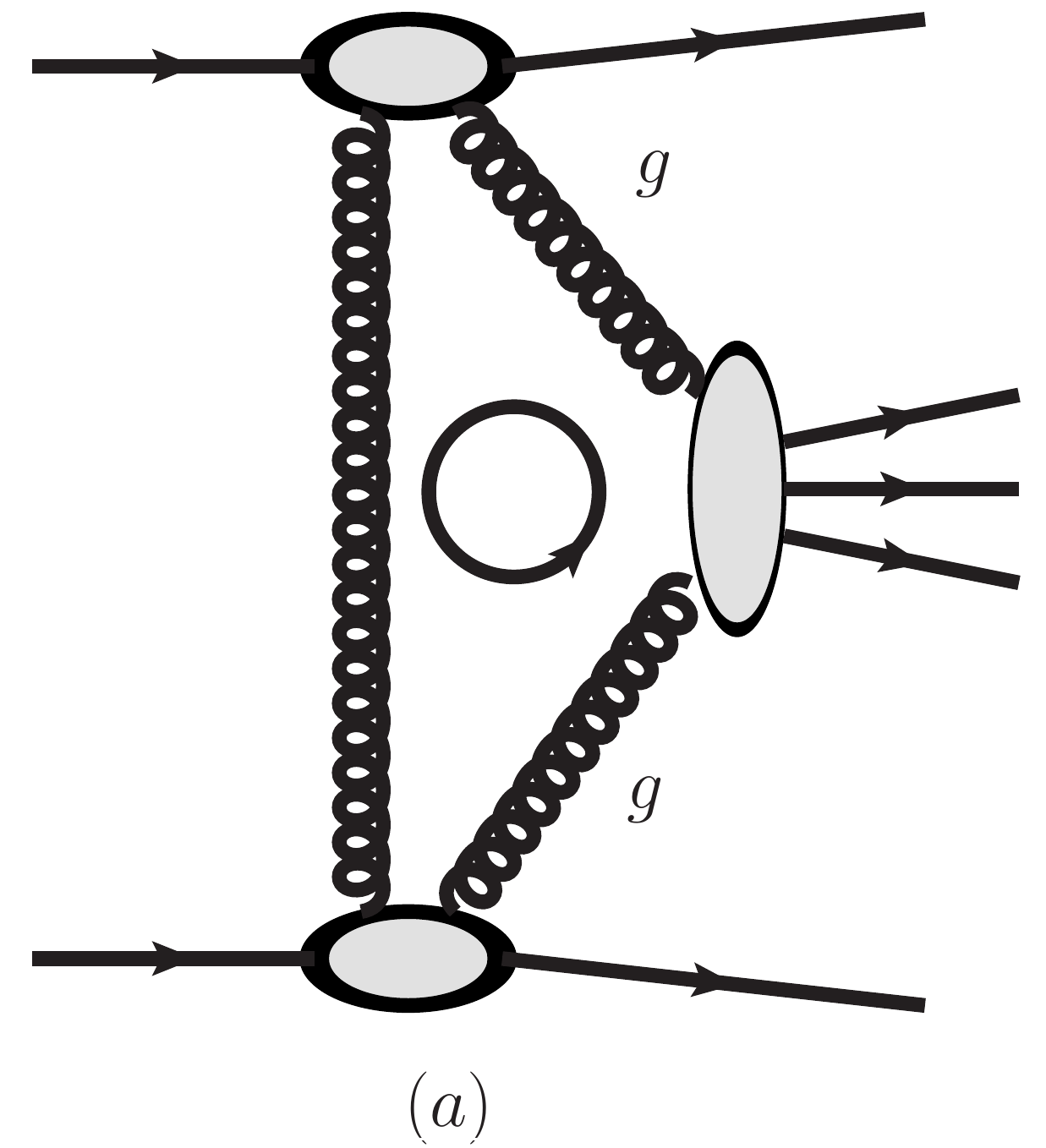}
\includegraphics[height=3.4cm,width=3.4cm]{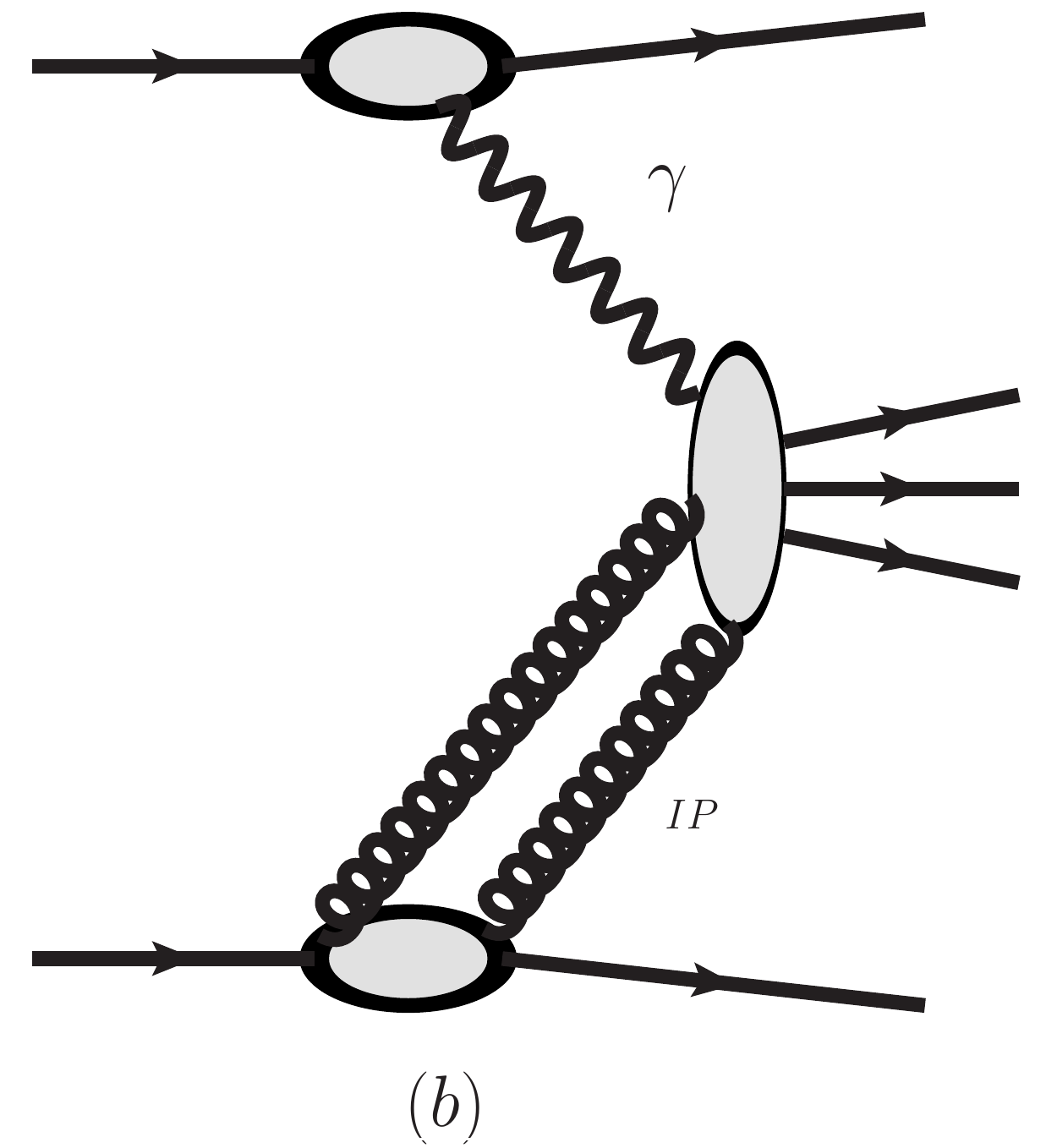}
\includegraphics[height=3.4cm,width=3.4cm]{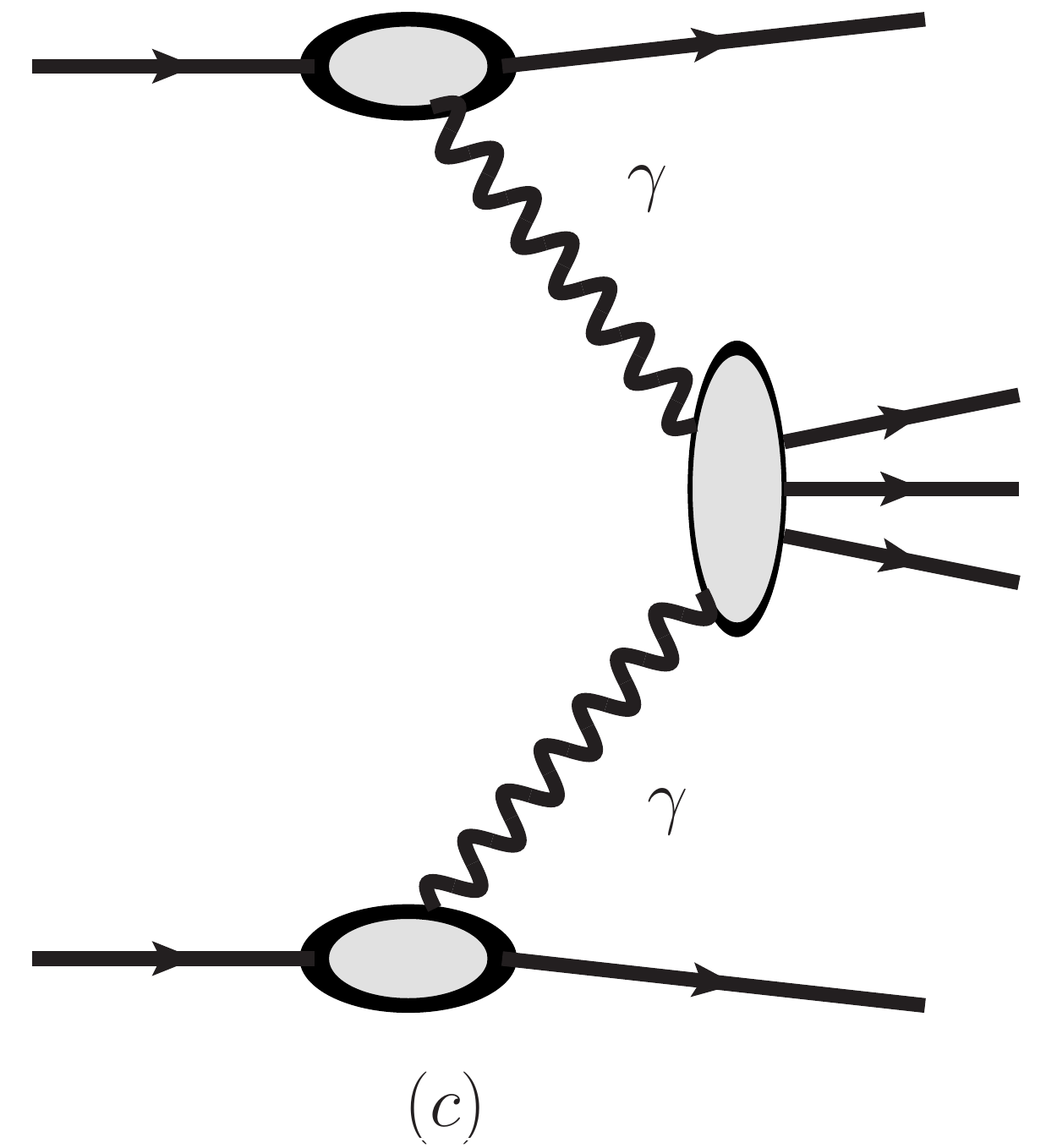}
\includegraphics[height=3.4cm,width=3.4cm]{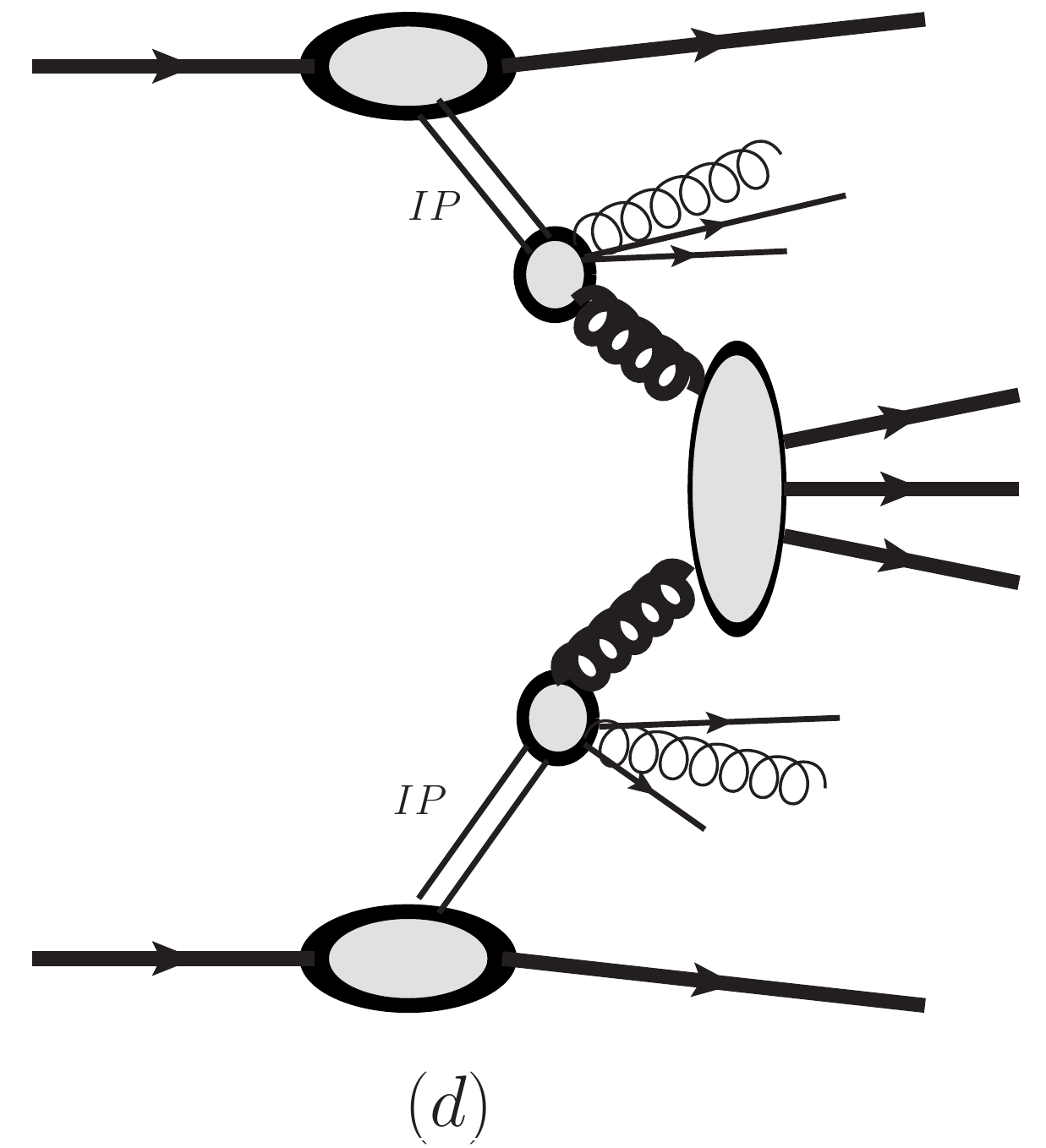}\includegraphics[height=3.4cm,width=3.4cm]{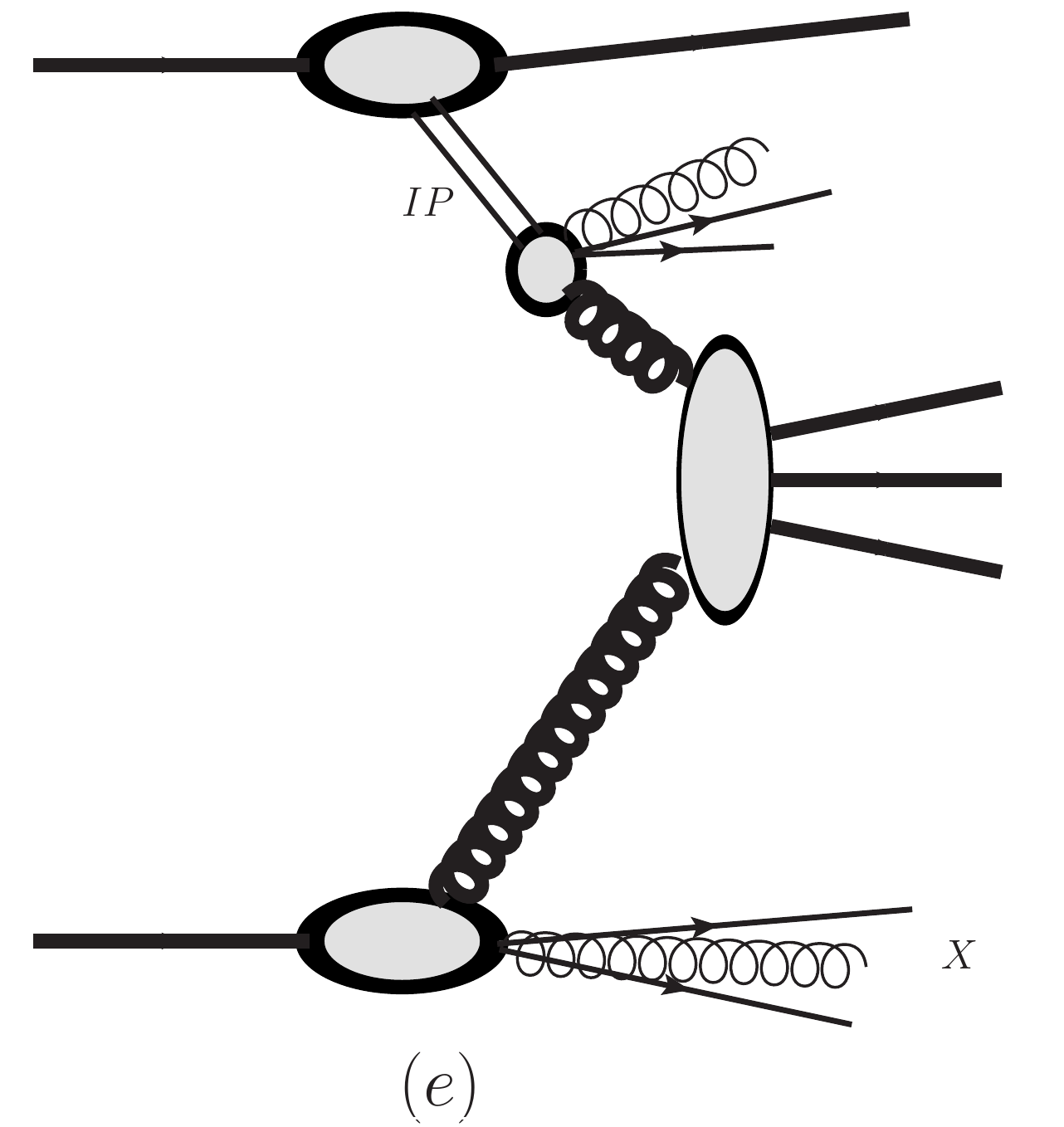}
%\end{minipage}
\caption{ \normalsize 
The illustrative description of  exclusive and inclusive processes.}
\label{fig1:limits}
\end{figure}
On top of that, it can be split into exclusive and inclusive event as displayed in Fig.\ref{fig1:limits}.  Fig.\ref{fig1:limits}(a) describes schematically the exclusive diffraction via two-gluon exchange between the two incoming protons. The soft pomeron is seen as a pair of gluons non-perturbatively coupled to the proton. One of the gluons is then coupled perturbatively to the hard process while the other one plays the role of a soft screening of colour, allowing for diffraction to occur\cite{Royon:2006kn}. The exclusive diffraction events comprise the presence of the rapidity gaps which separate the very intact forward outgoing protons from the centrally measured object produced alone. The intact forward outgoing protons are detected by forward hadron tagging detectors installed at low scattering angle with respect to beam axis near central detector \cite{Albrow:2008pn,Trzebinski:2015bra} after the exclusive difractive collisions. The whole Pomeron energy is used to produce the diffractive state, i.e there is no energy loss and Pomeron remnants\cite{Royon:2018hvi}. The leading protons carry most of the beam particle momentum and the full energy available is used in the interaction. The exclusive diffractive $\rm \chi_{cJ}$, $\rm \eta_c$, $\rm J/\psi$ productions and so on have been studied in Durham model with tagged proton or antiproton \cite{Harland-Lang:2014lxa,Pasechnik:2009bq,Pasechnik:2007hm,Khoze:2004yb,Pasechnik:2009qc,HarlandLang:2009qe,N.Cartiglia:2015gve} along with dedicated Monte Carlo codes \cite{Harland-Lang:2015cta}. The Figs.\ref{fig1:limits}(b,c) describe other exclusive productions such as photon-Pomeron and photon-photon fusion which are also very interesting. Notice that these purely exclusive production estimates can be useful to study the characteristics of the produced bound states or particles. However, their precise determinations lie on the consideration of the inclusive production\cite{Boonekamp:2002vg} which serves as {their important background. Here in our present paper we are concentrating on the inclusive diffractive production shown in Fig.\ref{fig1:limits}(d, e).

As for the inclusive diffractive processes, they differ from their counterparts by smaller rapidity gaps, and the colliding Pomerons or Reggeons are composite systems made from quarks and gluons. There are also presence of Pomeron or Reggeon remnants accompanied with soft QCD radiations. The presence of one intact forward hadron tagged in the final state and one large rapidity gap in the detector is called the single diffractive dissociation (SD). The central diffractive dissociation, double Pomeron or Reggeon exchange or Pomeron-Reggeon cross exchange (DD), is characterized by two intact forward hadrons and two rapidies. The experimental diffractive studies\cite{White:2008zzg,Abachi:1994hb,Abe:1994de,Aaltonen:2010qe,Aaltonen:2012tha,Affolder:1999hm} have particularly drawn attention toward the understanding of diffractive production due to measurement data samples released by the LHC. Theoretically, the inclusive diffractive processes have been studied in Regge theory (also named resolved-Pomeron model)
in Refs\cite{Ingelman:1984ns,Goncalves:2017bmo,Royon:2006by,Luszczak:2016csq} and so on. Taking into consideration the perturbative QCD and the soft diffractive physics, some studies have shown and addressed the non negligible Reggeon contribution \cite{Luszczak:2014mta,Luszczak:2014cxa}. In addition to the resolved-Pomeron model, existing models such as Donnachie-Landshoff \cite{Kochelev:1999wv,Donnachie:1987gu,Royon:2006kn} model and Bialas-Landshoff model\cite{Rangel:2006mm} have also investigated diffractive production of particles.

In this paper, we have predicted the cross section for the  inclusive single and double diffractive hadroproduction of $\rm \eta_{c}$ in proton-proton interactions in collinear momentum space in NRQCD formalism with Regge theory. The non diffractive hadroproduction (ND) has been estimated alongside diffractive ones. As said, on one hand such study	can be the background of exclusive production which requires precise determination, on the other hand, they themselves are also sensitive to gluon content of Pomeron (Reggeon) whereas the Pomeron (Reggeon) themselves are sensitive to the gluon distribution in the proton. Thus this kind of diffractive interaction is worth to be studied. Typically, we have also added Reggeon-Reggeon and Reggeon-Pomeron contributions which have usually ignored in former other calculations. We have concentrated on the $\rm \eta_{c}$ particle which is pseudoscalar particle of even charge parity. In this case, the dominant production mechanism is via the $\rm gg \to \eta_{c}$ gluon-gluon fusion 2-1 process at the pole $\rm z=1$, where z is defined as  $\rm z= P_{h}\cdot P^{\eta_{c}}/P_{h}\cdot P_{g}$ where $\rm P_{h}$, $\rm P_{g}$, $\rm P^{\eta_{c}}$ are the four momenta of proton, gluon and $\rm \eta_{c}$ respectively. The related diagrams are illustrated in Fig.\ref{fig2:limits}. Notice that in the standard collinear factorization approach there is zero transverse momentum distribution for the final $\rm \eta_{c}$, whereas the $\rm \eta_{c}$ production does possess transverse momentum distributions in the LHCb collaboration measurement. We therefore comment that it may also be interesting to consider, for example, the $\rm gg \to \eta_{c}+g$, $2\to 2$ processes and study the kinematical region $\rm 0<\rm z<1$ and even go beyond the leading order approximation up to NLO to include the full region. In our future work, the transverse momentum will be also included in the perturbative QCD and the soft diffractive parts. Anyway, this paper is our first step towards future work which is in progress. The $\rm \eta_{b}$ production is also included in our work. 
\begin{figure}[htp]
\centering
%\begin{minipage}[t]{4.0cm}
\includegraphics[height=3.8cm,width=5.4cm]{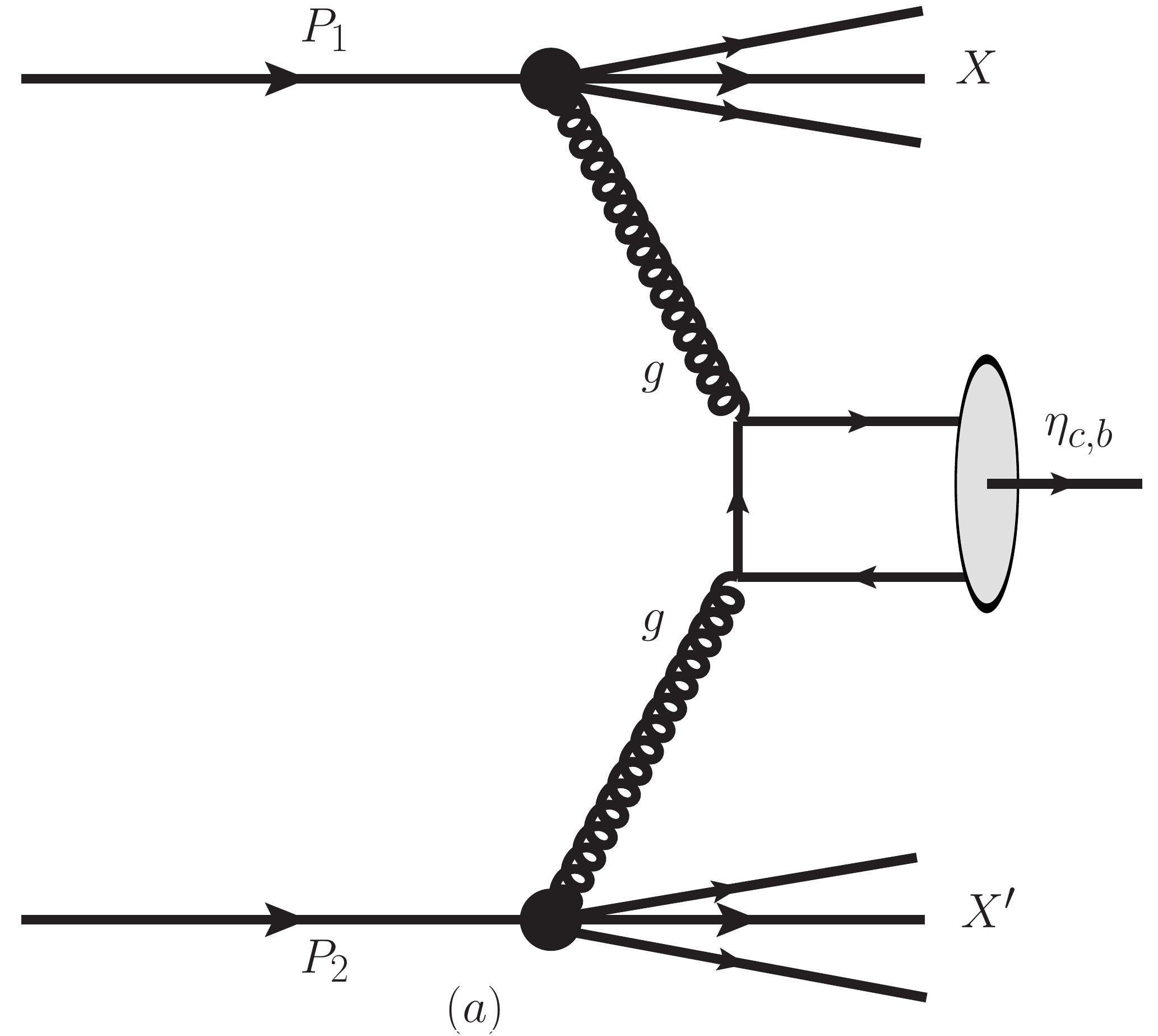}
\includegraphics[height=3.8cm,width=5.4cm]{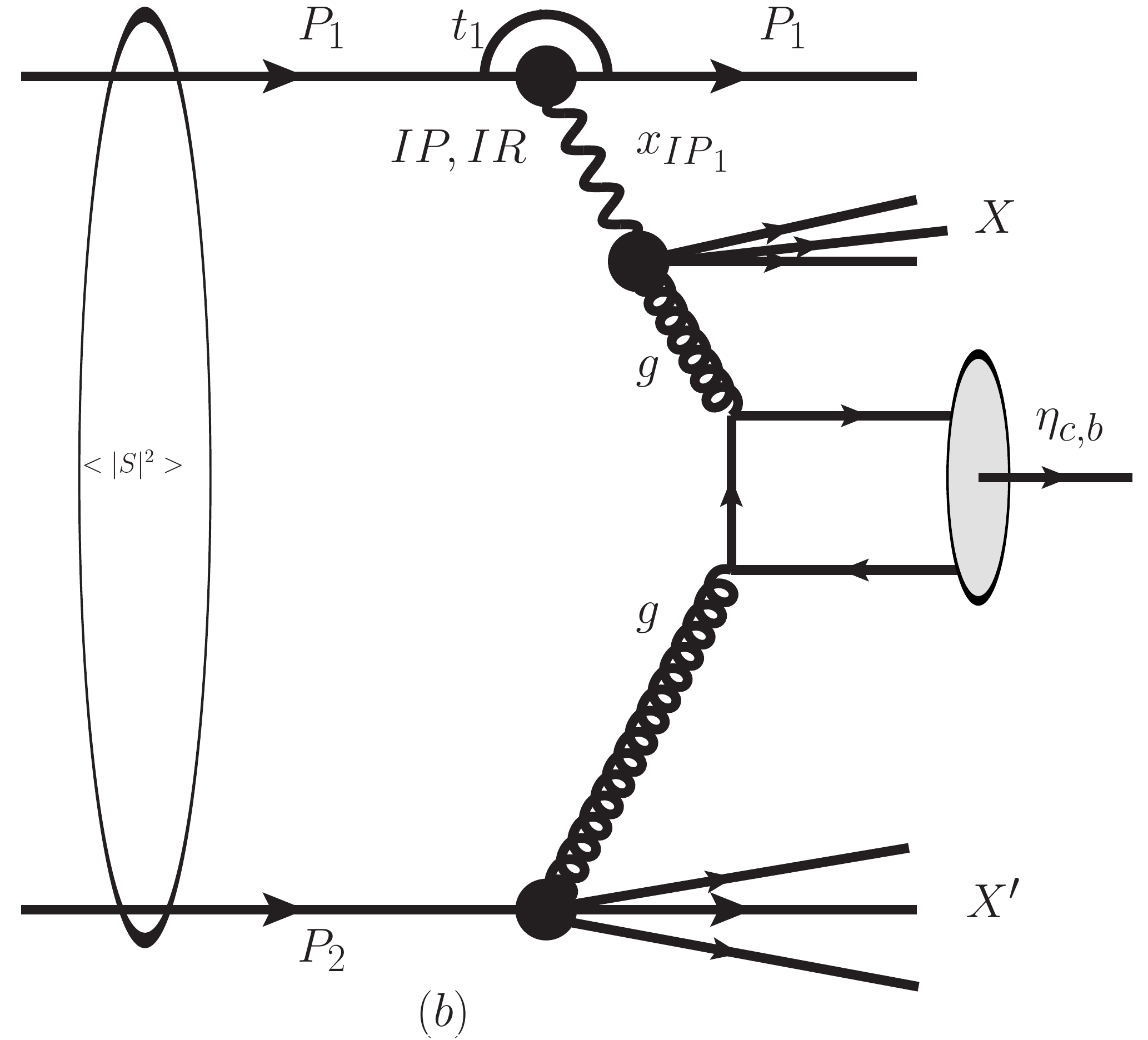}
\includegraphics[height=3.8cm,width=5.4cm]{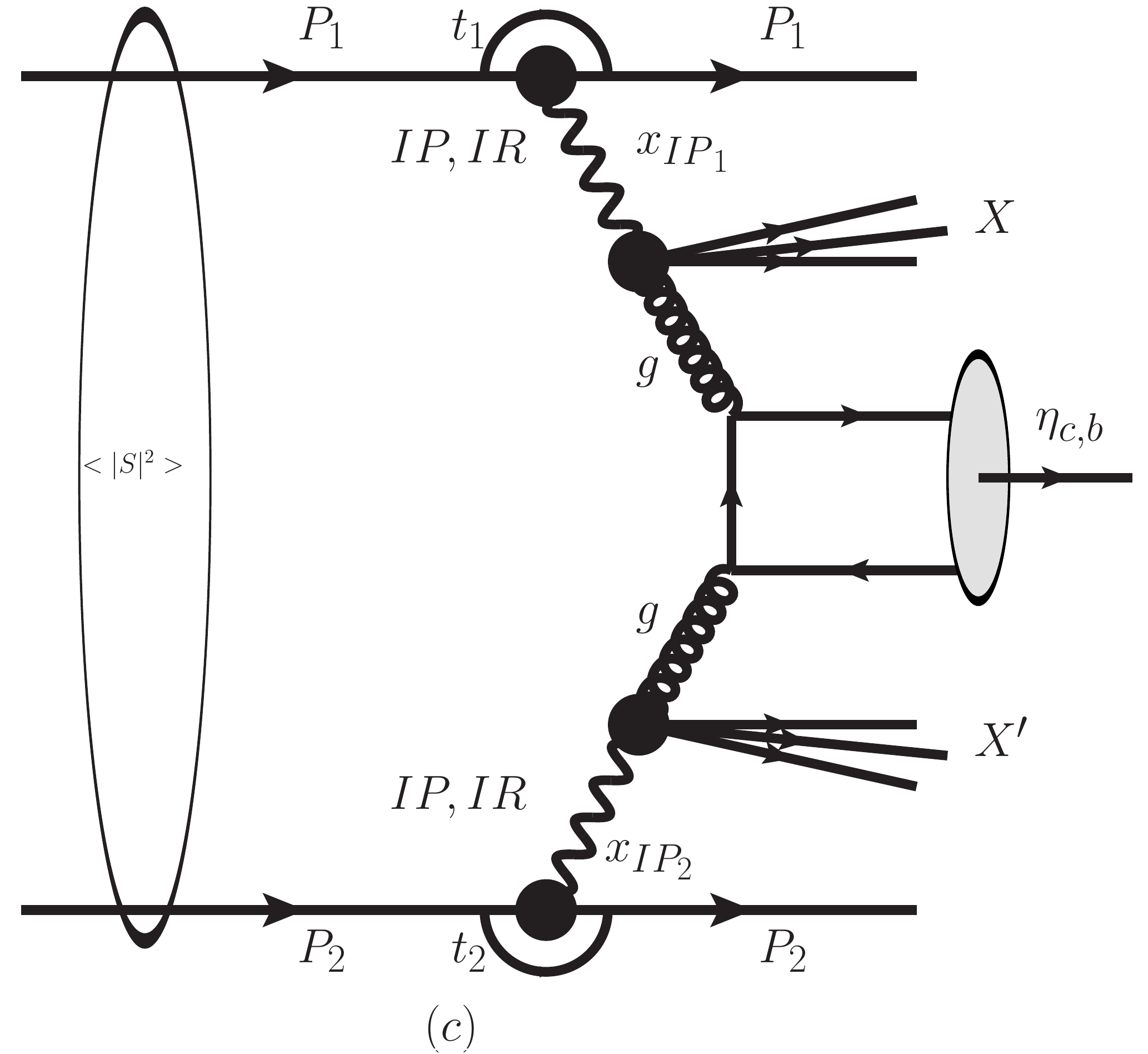}
%\end{minipage}
\caption{ \normalsize
Diagram representing single $\rm \eta_{c, b}$ quarkonium hadroproduction in non diiffractive (ND) (a), single diffraction SD(b) and double diffractive DD (c).}
\label{fig2:limits}
\end{figure}

The paper is structured into three sections including the introduction in Section \ref{Intro}. The detailed description of the hadron tagging devices and the formalism framework for the leading order cross section of $\rm \eta_{c}$ and $\rm \eta_{b}$ hadroproductions at the LHC are clearly described in Section \ref{FRAMEWORK}. The input parameters and discussed numerical results are shown in Section \ref{NUMERICAL}. A summary is briefly given in Section \ref{Conclusion}.

\section{FRAMEWORK OF THE CALCULATION TECHNOLOGY}
\label{FRAMEWORK}

\subsection{Forward diffractive detectors}

In SD and DD dissociations in $\rm pp$ collision, the produced single $\rm \eta_{c}$ or $\rm \eta_{b}$ accompanied with a low-mass system are measured in the central detector. The initial incoming hadron, which is described by the different components of its wave function, can be absorbed during the diffractive scattering process. The final outgoing intact hadron can be just excited into a diffractive system and observed by forward detectors. The hadronic diffraction is where intact particles are excited into hadronic system with small invariant mass, much smaller than the collision energy. The experimental signature of this process is that all hadrons are produced at small angles. The central detectors can give the information to help the forward detector to identify diffractive and non-diffractive events \cite{Zhou:2018elb}. In  non diffractive events, the color charge is exchanged between the interacting hadrons while the color singlet is exchanged in diffractive events \cite{Staszewski:2014hua}. The signal is to measure the exclusively produced heavy bound state in the central detector and the background is due to inclusive diffractive processes. The Pomeron or Reggeon remnants and and QCD radiation are detected by the central detectors. There are two valuable detector characteristics of the diffractive processes, namely the existence of the intact initial hadron and large rapidity gap which goes together with it. The rapidity gap size and the location of them in the pseudorapidity phase-space can be used to determine the type of the diffractions \cite{Zhou:2017kpg}. These rapidity gaps in the forward or backward rapidity regions, connect directly to the soft part of the events, and therefore nonperturbative effects, on a long space-time scale.

The detectors possess the coverage necessary to measure forward rapidity gaps, $\rm \Delta y$.  Negative (positive) pseudorapidity or the large polar plane is referred to as left (right) side of the detector for $\rm y < 0 $ ($\rm y > 0$). The pseudorapidity is often used experimentally instead of rapidity, which they are equal in the limit of a massless particle. These detectors try to measure both the cross sections and the kinematic properties of diffractive events at the LHC energies. The areas cover the range where proton are either both observed at 420 m (symmetric tagging) \cite{Adriani:2008zz} or one is detected at 220 m and other at 420 m (asymmetric tagging). The forward tagging hadron detectors are characterized by their acceptance, resolution and ability to measure the time-of-flight from interaction point.

The transverse and longitudinal momenta defined relative to the beam axis, the azimuthal angle around the beam axis and the pseudorapidity defined in terms of the polar angle $\rm \theta$ with respect to the beam axis, are kinematic variables for the diffractive processes. The coordinates ($\rm r$, $\rm \phi$) are used in the plane transverse to the beam axis. The relationship between the observable $\rm \Delta y$ size and $\rm \xi_{X}$  is given as $\rm \Delta y \simeq-log( \xi_{X})$ and $\rm \Delta y \simeq log( \xi_{X} \xi_{X'})$ where $\rm \xi_{X}$ is function of the invariant mass of the whole diffractive final state, $\rm M_{X}=\sqrt{\xi_{X}s}$ for single diffractive, $\rm M_{XX'}=\sqrt{\xi_{X}\xi_{X'}ss'}$ for double diffractive and the center of momentum energy \cite{Abelev:2012sea}. For proton tagging at the LHC, we have adopted a region of  $0.0015<\xi_{1}<0.5$, $0.1<\xi_{2}<0.5$ for CMS-TOTEM forward detector, and $0.015<\xi_{3}<0.15$ for AFP-ATLAS forward detector \cite{Albrow:2008pn}.

\subsection{Cross Section Formulations}

In the following section, we refer to the heavy quarkonia as $\rm \eta_{c}$ and $\rm \eta_{b}$ whereas $\rm h_{1}h_{2}$ is symbolized by $\rm pp$. The non diffractive (ND), single diffractive(SD) and double diffractive(DD) hadron-hadron reactions are given as
\begin{eqnarray}\nonumber
&&\rm ND:\ \ \ h_{1}h_{2}\to \eta_{c,b}\ X \\ \nonumber
&&\rm SD:\ \ \ \ h_{1}h_{2}\to h_{1}\otimes X+\eta_{c,b}+X{'} \\
&&\rm DD:\ \ \ h_{1}h_{2}\to h_{1}\otimes X+\eta_{c,b}+X{'}\otimes h_{2} .
\end{eqnarray}
The total cross section of non diffractive  is given by the convolution of partonic cross section and gluon distribution functions of the incident particles for the correspondent process in gluon-gluon fusion and can be written as
\begin{eqnarray}\label{a}\nonumber
\rm \sigma^{ND} (\rm h_{1}h_{2}\to \eta_{c,b}X)&=& \rm  \int_{0}^{1}\frac{dx_{1 }}{
x_{1}}\int_{0}^{1}\frac{dx_{2}}{x_{2}}\sum_{n}\hat{\sigma}(gg\rightarrow Q\overline{Q}
[n]+X) \\
&&\rm \langle 0|\mathcal{O}_{1}^{\eta_{c,b}}[n]|0\rangle\rm \lbrack \mathcal{F}_{g }(x_{1},\mu ^{2})\mathcal{F}_{g}(x_{2},\mu ^{2})+(h_{1}\leftrightarrow h_{2})],
\end{eqnarray}
for single diffractive process, the total cross section is written as
\begin{eqnarray}\label{b}\nonumber
\rm \sigma^{SD} (\rm h_{1}h_{2}\to h_{1}\otimes X +\eta_{c,b}+X^{'})&=&
 \rm \langle \left\vert S\right\vert ^{2}\rangle \int_{0}^{1}\frac{dx_{1 }}{%
x_{1}}\int_{0}^{1}\frac{dx_{2}}{x_{2}}\sum_{n}\hat{\sigma} (gg\rightarrow Q\overline{Q}%
[n]+X) \\
&&\rm \langle 0|\mathcal{O}_{1}^{\eta_{c,b}}[n]|0\rangle\rm \lbrack\mathcal{F}_{g}^{D}(x_{1},\mu ^{2})\mathcal{F}_{g}(x_{2},\mu ^{2})+(h_{1}\leftrightarrow h_{2})],
\end{eqnarray}
as for double diffractive process, the total cross section is formulated as
\begin{eqnarray}\label{b}\nonumber
\rm \sigma^{DD} (\rm h_{1}h_{2}\to h_{1}\otimes X +\eta_{c,b}+h_{2}\otimes X^{'})&=&
 \rm \langle \left\vert S\right\vert ^{2}\rangle \int_{0}^{1}\frac{dx_{1 }}{%
x_{1}}\int_{0}^{1}\frac{dx_{2}}{x_{2}}\sum_{n}\hat{\sigma} (gg\rightarrow Q\overline{Q}%
[n]+X) \\
&&\rm \langle 0|\mathcal{O}_{1}^{\eta_{c,b}}[n]|0\rangle\rm \lbrack\mathcal{F}_{g}^{D}(x_{1},\mu ^{2})\mathcal{F}_{g}^{D}(x_{2},\mu ^{2})+(h_{1}\leftrightarrow h_{2})],
\end{eqnarray}
where $\rm \langle 0|\mathcal{O}_{1}^{\eta_{c,b} }[n]|0\rangle$ is the long-distance matrix element which describes the hadronization of the $\rm Q\overline{Q}$ heavy pair into the physical observable quarkonium state $\rm \eta_{c}$ or $\rm \eta_{b}$.
The $\rm \hat{\sigma}(gg\rightarrow Q\overline{Q}[n])$ denotes the short-distance cross sections for the partonic process $\rm gg\rightarrow Q\overline{Q}[n]$, which is found by operating the covariant projection method \cite{Tichouk:2019lxk,Petrelli:1997ge}.
The Fock state $\rm n$ are given as follows:  $\rm ^{1}S_{0}^{[1]}$, $\rm ^{1}S_{0}^{[8]}$ for $\rm gg\to Q\overline{Q}[n]$ partonic process \cite{Basu:2002rk}. The contribution of color singlet states for $\rm \eta_{c}$ and $\rm \eta_{b}$ quarkonium production is at leading power in velocity (v) while the color octet contribution to S-wave quarkonium production are power suppressed \cite{Kang:2013hta}. The $\rm \mathcal{F}_{g }(x_{i},\mu ^{2})$ and $\rm \mathcal{F}_{g}^{D}(x_{i},\mu ^{2})$ stand for the conventional integrated gluon parton distribution function (PDF) in the proton and their diffractive counterparts, respectively. $\rm x_{i}$ is Bjorken variable defined as the momentum fractions of the hadron (proton), momentum carried by the gluons. $\rm \langle \left\vert S\right\vert ^{2}\rangle$ is the gap survival probability or total factor. The partonic cross section is
\begin{eqnarray}
\rm \hat{\sigma}(gg\rightarrow Q\overline{Q}[n])=\frac{\pi}{M_{\eta_{c,b}}^{2}}\delta(\hat{s}-M_{\eta_{c,b}}^{2})\overline{\sum }|\mathcal{A}_{S,L}|^{2}
\end{eqnarray}{}
with the matrix element squared is given by \cite{PhysRevD.88.014008,Maltoni:2004hv,Cho:1995ce}
\begin{eqnarray}
\rm \overline{\sum } |\mathcal{A}_{S,L}|^{2}=\frac{2}{9}\frac{\pi^2\alpha^{2}_{s}}{M_{\eta_{c,b}}}\langle 0|\mathcal{O}_{1}^{\eta_{c,b}}(^{1}S_{0})|0\rangle + \frac{5}{12}\frac{\pi^2\alpha^{2}_{s}}{M_{\eta_{c,b}}}\langle 0|\mathcal{O}_{8}^{\eta_{c,b}}(^{1}S_{0})|0\rangle
\end{eqnarray}{}
and the colliding energie is written as  $\rm \hat{s}=\rm x_{1}x_{2}s$.

\subsection{Gap survival probability in diffractive processes}

The gap survival probability \cite{Khoze:2017sdd} is characterized by the presence of additional soft partonic interactions and new particles in gap rapidity. It can be described by additional soft incoming or outgoing proton-proton rescatterings with multi-Pomerons exchanged (eikonal factor), by the interaction of incoming or outgoing proton with intermediate partons (enhanced factor), by gluon radiation from annihilation of two energetic coloured particles called hard QCD bremsstrahlung (Sudakov factor) and by the change of the forward intact proton momentum (migration). 

The enhanced and Sudakov factors as well as the migration are neglected in collinear approximation where the transverse momenta of intermediate partons and  screening gluon are not taken into consideration and,
the incoming proton and outgoing intact forward proton have almost the same direction. The eikonal gap survival probability has been evaluated as $\rm \langle \left\vert S\right\vert ^{2}\rangle $    \cite{Luszczak:2014mta,Sun:2017jcg,Muller:1990sg} and reads
\begin{eqnarray}
\rm \langle \left\vert S\right\vert ^{2}\rangle_{pp} =   \frac{B_{1}}{B_{\mathbb{P}}}(\frac{\sigma^{tot}_{pp}(s)}{2\pi B_{1}})^{-B_{1}/B_{\mathbb{P}}}\gamma(B_{1}/B_{\mathbb{P}},\frac{\sigma^{tot}_{pp}(s)}{2\pi B_{1}})
\end{eqnarray}
where $\gamma$ is the incomplete gamma function, $\sqrt{s}$=13 TeV, $\rm B_{1}=\frac{B_{0}}{2}+\frac{\alpha'}{2}ln(\frac{s}{s_{0}})$ \cite{Guzey:2016tek,Basso:2017mue}, $\rm s_{0}=1$ GeV$^{2}$ for two channel model \cite{Gotsman:2005wa}, B$_{0}$=10\ GeV$^{-2}$, $\alpha$'=0.25 GeV$^{-2}$, $\rm B_{2}=\frac{1}{2Q^{2}_{0}}+\frac{B_{1}}{4}$. The chosen total pp cross section is parameterized by the optical theorem in one ways as  
$\rm \sigma^{tot}_{pp}(s)= 33.73+0.2838\ln^{2}(s)+13.67s^{-0.412}-7.77s^{-0.5626}$ mb
\cite{Agashe:2014kda,Klein:2016yzr} and in other way as $\rm \sigma^{tot}_{pp}(s)= 69.3286+12.6800\ln(\sqrt{s})+1.2273\ln^{2}(\sqrt{s})$\cite{Kohara:2014waa}. Thus its computed value is $\rm \langle \left\vert S\right\vert^{2}\rangle_{pp} = 0.09 (0.03)$ for the LHC energy. The approximative formula is also given as \cite{Luszczak:2014mta,Sun:2017jcg,Muller:1990sg}
\begin{eqnarray}
&&\rm \rm \langle \left\vert S\right\vert ^{2}\rangle_{pp} = \frac{a}{b+\ln(\sqrt{s/s_{0}})}, 
\end{eqnarray}
with $\rm a = 0.126$, $\rm b=-4.688$ and this approximative value is $\rm \langle \left\vert S\right\vert ^{2}\rangle_{pp} = 0.03$ \cite{Marquet:2016ulz}. Those additional soft interactions from eikonal factor can destroy the diffractive signature \cite{Bjorken:1992er} and the Regge factorization is known to be violated in the treatment of diffractive interactions in hadronic collisions. The gap survival probability
relies on the specific collision, the cuts prescribed in the experiment and stands for the last element of the resolved-Pomeron model. A variety of attempts have been carried out to estimate those probabilities \cite{Kopeliovich:2016rts,Gotsman:2011xc,Khoze:2008cx}, however the actual values are rather uncertain. The selected value can be regarded  a lower limit given the recent available experimental results \cite{Chatrchyan:2012vc,Aad:2015xis}.

\subsection{The Pomeron and Reggeon parton distribution functions}

As stated by the so-called proton-vertex factorization or the Resolved Pomeron Model \cite{Ingelman:1984ns}, the collinear diffractive gluon, $\rm g_{p}^{D}(x_{g}, \mu^{2}_{f}, x_{\mathbb{P}})$ is defined as a convolution of the Pomeron (Reggeon) flux emitted by the proton, $\rm f^{h}_{\mathbb{P,R} } (x_{\mathbb{P}} )$, and the gluon distribution in the Pomeron (Reggeon), $\rm g^{\mathbb{P,R}}(\beta,Q^{2})$  where $\beta(=\frac{x_{1}}{x_{\mathbb{P}}})$ is the longitudinal momentum fraction carried by the partons inside the Pomeron. The Reggeon contribution is ignored in the hard diffraction calculations of different final states in most cases. The Reggeon contribution is treated as an exchange of quark and antiquark pair and the parton content of the Reggeon is obtained from the pion structure function \cite{Aktas:2006hy}. The difference between the two contributions exists in the $\rm x_{\mathbb{P}}$ and $\rm t$ dependence of their fluxes, where the Reggeon exchange is mostly significant at high $\rm x_{\mathbb{P}}$, remarkably for $\rm x_{\mathbb{P}}>0.1$. $\rm x_{\mathbb{P}}$ stands also for $\xi$. The Reggeon shape of the $\rm t$ distribution is also different showing a less steep decrease than in the Pomeron case. Nevertheless, as shown in Ref.\cite{Marquet:2016ulz}, this contribution is significant in some regions of the phase space and needed to obtain a good description of the data, the collinear diffractive gluon distribution of the proton at low $\rm \beta$ and large $\rm x_{\mathbb{P}}$ \cite{Rasmussen:2015qgr,Aktas:2006hx} is formulated by
\begin{eqnarray}\label{654}
\mathrm{g_{p}^{D}(x_{g},Q^{2},x_{\mathbb{P}})=\int_{x_{g}}^{1}\frac{dx_{\mathbb{P}}}{x_{\mathbb{P}}}
f_{\mathbb{P}}^{p}(x_{\mathbb{P}})g^{\mathbb{P}}(\frac{x_{g}}{x_{\mathbb{P}}},Q^{2})+n_{\mathrm{\mathbb{R}}
}\int_{x_{g}}^{1}\frac{dx_{\mathbb{P}}}{x_{\mathbb{P}}}f_{\mathbb{R}}^{p}(x_{\mathbb{P}})g^{R}(\frac{x_{g}}{
x_{\mathbb{P}}},Q^{2})}
\end{eqnarray}
and the Pomeron and Reggeon fluxes are literally expressed by
\begin{eqnarray}\label{m}
\rm  f_{\mathbb{P,R}}^{p}(x_{\mathbb{P}})= & \rm
\int_{t_{\min }}^{t_{\max
}}dtf_{\mathbb{P},\mathbb{R}/p}(x_{\mathbb{P}},t)= & \rm
\int_{t_{\min }}^{t_{\max }}dt\frac{A_{\mathbb{P,R}}e^{B_{\mathbb{P,R}}t}
}{x_{\mathbb{P}}^{2\alpha _{\mathbb{P,R}}(t)-1}},
\end{eqnarray}
where the variables, $\rm t_{\min}$ and $\rm t_{\max}$, are kinematically fixed limits. The Pomeron (Reggeon) flux factor is stimulated by Regge theory, where the Pomeron(Reggeon) trajectory is linearly supposed to be, $\rm \alpha_{\mathbb{P,R}}(t)= \alpha_{\mathbb{P,R}}(0)+\alpha_{\mathbb{P,R}}^{\prime}t$, and the parameters $\rm B_{\mathbb{P,R}}$ , $\rm \alpha_{\mathbb{P,R}}^{\prime}(t)$ and their uncertainties are taken from fits to H1 data \cite{Aktas:2006hy}. The slope of the Pomeron(Reggeon) flux is $\rm B_{\mathbb{P,R}}=5.5^{-2.0}_{+0.7}(1.6^{-1.6}_{+0.4})$ GeV$^{-2}$, the Regge trajectory of the Pomeron(Reggeon) $\rm \alpha_{\mathbb{P,R}}(t)= \alpha_{\mathbb{P,R}}(0)+\alpha^{\prime}_{\mathbb{P,R}}(t)$ with $\rm  \alpha_{\mathbb{P,R}}(0)=1.118\pm 0.008(0.50\pm0.10)$ and $\rm \alpha^{\prime}_{\mathbb{P,R}}=0.06^{+0.19}_{-0.06}$ GeV$^{-2}(0.3^{+0.6}_{-0.3}$ GeV$^{-2})$. The $\rm t$ integration limits are $\rm t_{max} = -m^{2}_{p}x^{2}_{ \mathbb{P} }/(1-x_{\mathbb{P}} )$ ($\rm m_{p}=0.93827231$ GeV symbolizes the proton mass) and $\rm t_{min} = -1$ GeV$^{2}$. Lastly, the normalization factor $\rm A_{\mathbb{P,R}} = 1.7101(1705.0)$ is selected such that $\rm x_{\mathbb{P}}\times \int_{t_{\min }}^{t_{\max }}dt f_{\mathbb{P},\mathbb{R}/p}(x_{\mathbb{P}},t)=1$ at $\rm x_{\mathbb{P}} = 0.003$ and $\rm n_{\mathrm{\mathbb{R}}}=(1.7\pm 0.4)\times10^{-3}$. The $\rm f^{p}_{\mathbb{P,R}} (x_{\mathbb{P}} )$ is the Pomeron (Reggeon) flux factor which describes the emission rate of Pomeron (Reggeon) by the hadron ($\rm p$) and represents the probability that a Pomeron with particular values of ($\rm x_{\mathbb{P}}; t $) couples to the hadron. Certain fraction of the pomeron energy is only available for the hard collision and the rest being carried away by a remnant or spectator jet. On every occasion a coloured parton (gluon) is pulled out of a colour-singlet object, pomeron. The Pomeron structure is well restricted by the fits, fit A and fit B, which evidently reveal that its parton content is gluon dominated. Contrariwise, the HERA data do not restrain the parton distribution function of Reggeon which is therefore needed in order to get a quantitative description of the high-$x_{\mathbb{P}}$ measurements. Consequently, measurements at the LHC will permit to examine the validity of this supposition.

\section{NUMERICAL RESULTS AND DISCUSSION}
\label{NUMERICAL}

In the following part, we discuss the numerical results of the inclusive and diffractive hadroproduction of $\rm \eta_{c,b}$ by using some physical parameters such as: The masses of the heavy quarks are chosen as $\rm m_{c}$=1.45 GeV and $\rm m_{b}$=4.75 GeV. The mass of $\rm \eta_{c,b}$ is literally put at $\rm M_{\eta_{c,b}}=2m_{c,b}$. The colliding energy used in this paper is $\sqrt{s}$=13 TeV for $\rm pp$. The unpolarized distribution function of gluon from the proton, we adopt the leading-order set of the MSTW2008 parametrization \cite{Martin:2009iq}. The 2006 H1 proton diffractive PDFs (fit A) is used for the pomeron densities inside the proton\cite{Aktas:2006hy,Aktas:2006hx} which are probed at the factorization hard scale ($\mu=Q$) chosen as $\rm \mu=m^{\eta_{c,b}}_{T}$, where $\rm m^{\eta_{c,b}}_{T}=M_{\eta_{c,b}}$ is the $\rm \eta_{c,b}$ transverse mass. Numerical calculations are carried out by in-house monte carlo generator. The choice of the LDMEs for $\rm \eta_{c,b}$ is taken from \cite{Butenschoen:2011yh,Maltoni:2004hv,Yu:2017pot} and valued to $\rm \langle 0|\mathcal{O}_{1}^{\eta_{c}}(^{1}S_{0})|0\rangle$ =0.44 GeV$^3$, $\rm \langle 0|\mathcal{O}_{8}^{\eta_{c}}(^{1}S_{0})|0\rangle$ =0.00056  GeV$^3$, $\rm \langle 0|\mathcal{O}_{1}^{\eta_{b}}(^{1}S_{0})|0\rangle$ =3.63333  GeV$^3$ and $\rm \langle 0|\mathcal{O}_{8}^{\eta_{b}}(^{1}S_{0})|0\rangle$ = 0.0159 GeV$^3$.

\subsection{The Cross Sections}

In Table \ref{tab:1a}, the total cross section predictions of $\eta_{c}$ hadroproduction in ND, SD and DD processes are displayed for three different forward detector acceptances at the distinct ranges, $0.0015<\xi_{1}<0.5$, $0.1<\xi_{2}<0.5$ and $0.015<\xi_{3}<0.15$. In the single diffraction dissociation, we have noticed that the $\mathbb{R}$p contributions are substantial than that of $\mathbb{P}$p contributions for $\xi_{1,2}$ whereas the large contribution to the total cross section for $\xi_{3}$ hails from the $\mathbb{P}$p interactions. That means that the Reggeon contribution should be taken into consideration for some Reggeon longitudinal momentum fraction ranges at the LHC experiments, and should not be neglected particularly for $\xi_{1,2}$. As for the double diffraction dissociation, $\mathbb{RR}$ and $\mathbb{PR+PR}$ interactions provide more contributions to the total cross section of $\eta_{c}$ for $\xi_{1,2}$. In the case of $\xi_{3}$, the large contribution to the total cross section comes from the $\mathbb{PR+PR}$ cross exchange interactions. Reggeon contributions are still yet important for some same forward detector acceptances like in SD process. The $\mathbb{RP}$ and $\mathbb{RR}$ interactions should play a non-negligible contribution to the $\eta_{c}$ hadroproduction for Pomeron/Reggeon longitudinal momentum fraction range, as far as the $\xi_{1,2}$ are concerned. We have also seen, the non diffractive prediction is a factor 10$^{2}$ (10$^{3}$), 10$^{2}$ (10$^{3}$) and 10$^{3}$ (10$^{4}$)  larger
than of SD (DD) prediction for $0.0015<\xi_{1}<0.5$, $0.1<\xi_{2}<0.5$ and $0.015<\xi_{3}<0.15$, respectively. The total cross sections which made the approximation of neglecting the Reggeon contributions even more problematical. Reggeon can be more contributing that the Pomeron to total quarkonium  cross section in some kinematical ranges where it clearly dominates, and Reggeon can be experimentally isolated.
\begin{table*}[htbp]
\begin{center}
\begin{tabular}{|c | c | c | c | c | }
\hline
\multicolumn{2}{|c|}{\diagbox{Process}{$\xi_{i}$}}  &$0.0015<\xi_{1}<0.5$&$0.1<\xi_{2}<0.5$&$0.015<\xi_{3}<0.15$\\
\hline
   &$ \mathbb{P}$p  & 11.8  & 3.84 &5.72 \\
SD &$ \mathbb{R}$p  & 20.6  &18.6  &3.72 \\
   & total          & 32.4  & 22.4 &9.44 \\
\hline
   & $\mathbb{PP}$   &4.74$\times$10$^{-1}$  &1.74$\times$10$^{-2}$ &9.32$\times$10$^{-2}$ \\
DD &$\mathbb{PR+RP}$ &2.25  &6.14$\times$10$^{-1}$ &2.16$\times $10$^{-1}$ \\
   &  $\mathbb{RR}$  &1.79  &1.45 &6.06$\times$10$^{-2}$ \\
   & total           & 4.51 &2.08 &3.70$\times10^{-1}$ \\
\hline	
\hline
ND &gg               & \multicolumn{3}{|c|}{2050.07} \\
\hline
\end{tabular}
\end{center}\caption{\label{tab:1a}
The total cross section ($\mu$b) for $\eta_{c}$ hadroproduction at the LHC with forward detector acceptances.}
\end{table*}

In Table \ref{tab:1b}, we have also estimated the $\eta_{b}$ cross section for ND, SD and DD dissociations for $\rm \xi_{1}$ only where the Reggeon contribution to cross section is little bit sizable. The non diffractive prediction is a factor 10$^{2}$ (10$^{3}$) larger than SD (DD) prediction of $\eta_{b}$ hadroproduction for $0.0015<\xi_{1}<0.5$. The non diffractive, single and double diffractive cross section of $\eta_{c}$ is a factor 10$^{2}$ larger than that of $\eta_{b}$ due to its small mass.
\begin{table*}[htbp]
\begin{center}
\begin{tabular}{|c |c | c |  }
\hline	
 \multicolumn{2}{|c|}{\diagbox{Process}{$\xi_{i}$}} & \multicolumn{1}{c|}{$0.0015<\xi_{1}<0.5$}
\\\hline
   &$ \mathbb{P}$p  & 1.66$\times$10$^{-1}$    \\
SD &$ \mathbb{R}$p  & 2.39$\times$10$^{-1}$  \\
   &total  &  4.05$\times$10$^{-1}$  \\
\hline
   &$\mathbb{PP}$   &9.31$\times$10$^{-3}$  \\
DD &$ \mathbb{PR+RP}$  &  2.82$\times$10$^{-2}$ \\
   &$ \mathbb{RR}$  & 1.82$\times$10$^{-2}$  \\
   &total  &  5.57$\times$10$^{-2}$  \\
\hline
\hline
ND &gg & 24.68
\\
\hline
\end{tabular}
\end{center}\caption{\label{tab:1b}
The total cross section ($\mu$b) for the $ \eta_{b}$ hadroproduction at LHC for forward detector acceptances, $0.0015<\xi_{1}<0.5$.}
\end{table*}

The diffractive production rates for $\eta_{c}$ and $\eta_{b}$ in $\rm pp$ interactions by assuming the design integrated luminosities $\rm \mathcal{L}_{LHC}^{pp}= 10^{4}\mu b^{-1}s^{-1}$ and run times ($\rm T= 10s$) \cite{Goncalves:2013oga}. The production rate in ND, SD and DD processes are more sensitive to proton momentum loss, $0.0015<\xi_{1}<0.5$, due to its considerable Reggeon event numbers. The non diffractive event rate keeps the same  order of magnitude larger than SD and DD  as for $\eta_{b}$ and $\eta_{c}$ cross section predictions. The LHC capabilities should be utilized in order to constrain it better, and improve the theoretical event rate predictions of the various Pomeron/Reggeon longitudinal momentum fraction range studies.

The estimate of  uncertainty in inclusive diffractive cross section predictions arises from many sources. Firstly, it can been evaluated from different choices of the heavy quark masses, the long distance elements, the factorization scale or renormalization scale \cite{Bi:2016vbt}. Secondly,  the uncertainty can come from the gap survival probability which gives maybe the largest uncertainty of about  $\pm 50$\% \cite{Kaidalov:2003fw} or around 30 \% \cite{Luszczak:2014mta} in the overall production rate. The obtained results can be multiplied by a factor of $\rm 4$ at CMS collider \cite{Chatrchyan:2012vc}.  Thirdly, the error can be computed by the choice of two different diffractive PDF fits, H1 2006 dPDF Fit A and H1 2006 dPDF Fit B. The results are found lightly different between these two fits. Fourthly, the gluon density at high $\beta$ is however poorly known and the uncertainty is of the order of 25 \%. This high  $\beta$ region is of particular interest for the LHC since it represents for example a direct background to the search for exclusive events\cite{Royon:2006jf}. This uncertainty  takes into account the uncertainty of QCD fits at high $\beta$ and is related only to the gluon density from Pomeron $\rm g^{\mathbb{P}}(\frac{x_{g}}{x_{\mathbb{P}}},Q^{2})$, which is multiplied by an uncertainty factor $\rm (1-\beta)^{\nu}$ with $\rm \nu = -0.5$ or 0.5 \cite{Kepka:2007nr,Royon:2006by}. It is evaluated in Table \ref{tab:1c} for total SD and DD cross section of $\eta_{c}$ and $\eta_{b}$ difractive hadroproductions in $\rm pp$ collisions for $0.0015<\xi_{1}<0.5$. Fifthly, it is also worth mentioning that uncertainty range of the Reggeon contribution is unknown in literature. Sixthly, there are uncertainties which hail from the infrared region where the gluon distribution is not well understood as well as the uncertainty in the gluon distribution itself. And finally, it has  been noted that in the higher-order QCD, radiative corrections cause additional uncertainties \cite{HarlandLang:2009qe}.

\begin{table*}[htbp]
	\begin{center}
		\begin{tabular}{|c |c | c | c |}
			\hline	
			\multicolumn{3}{|c|}{\diagbox{Process}{$\xi_{i}$}}  & $0.0015<\xi_{1}<0.5$ \\
			\hline
			& SD&$\mathbb{P}$p  &   [10.12 ; 15.46] \\
			\multirow{3}{*}{$\eta_c$}&   &total    &   [30.73 ; 36.07] \\
			\cline{2-4}
			&DD &$\mathbb{PP}$      & [4.10$\times10^{-1}$ ; 5.90$\times10^{-1}$] \\
			&   &$ \mathbb{PR+RP}$  & [1.94 ;  2.88 ] \\
			&   &total       & [4.14 ;  5.26 ]\\
			\hline
			\hline
			\multirow{3}{*}{$\eta_b$}& SD &$ \mathbb{P}$p  & [1.53$\times$10$^{-1}$ ; 1.87$\times$10$^{-1}$]   \\
			&           &total     & [3.92$\times10^{-1}$ ; 4.27$\times10^{-1}$] \\
			\cline{2-4}
			&DD&$ \mathbb{PP}$    & [8.18$\times10^{-3}$ ; 1.11$\times10^{-2}$] \\
			&  &$ \mathbb{PR+RP}$  & [2.56$\times$10$^{-2}$ ; 3.23$\times$10$^{-2}$]\\
			&  &total  & [5.21$\times$10$^{-2}$ ; 6.16$\times$10$^{-2}$]  \\
			\hline
		\end{tabular}
	\end{center}\caption{\label{tab:1c}
		The uncertainties for the $\eta_{c}$ and $\eta_{b}$ hadroproductions at LHC for forward detector acceptances, $0.0015<\xi_{1}<0.5$.}
\end{table*}

The predictions are influenced by large theoretical errors as mentioned above. Those uncertainties can be suitably lessened by taking into account the ratio $\rm R_{i}$ of diffractive to non diffractive cross sections and double diffractive to single diffractive cross,
\begin{eqnarray}
\rm  R_{1}&=&\rm \frac{\sigma_{SD}}{\sigma_{ND}};  \rm R_{2}=\rm \frac{\sigma_{DD}}{\sigma_{ND}}; \rm  R_{3}=\rm \frac{\sigma_{DD}}{\sigma_{SD}},
\end{eqnarray}
which give the advantage to reduce experimentally systematic errors. The ratios have been measured in a range of final states at the Tevatron and, certain stable behaviors with a value near 1\% has been displayed \cite{Affolder:1999hm, Aaltonen:2010qe,Aaltonen:2012tha}. We have presented these ratios of cross sections of $\eta_{c}$ and $\eta_{b}$ in Table \ref{tab:1d}. Our $\rm \eta_{c}$ ratio results have indicated that the single diffractive dissociation to non diffractive process provides the leading order estimate of 1.58\%, 1.09\% and 0.46\% for $0.0015<\xi_{1}<0.5$,$0.1<\xi_{2}<0.5$ and $0.015<\xi_{3}<0.15$, respectively. As for $\eta_{b}$, the ration is of 1.64\% for $0.0015<\xi_{1}<0.5$.
日\begin{table*}[htbp]
\begin{center}
\begin{tabular}{|c | c | c | c | c | }
\hline
\multirow{1}{*}{$\rm R_{i} \diagdown \xi_{i}$}  & \multicolumn{1}{c|}{$0.0015<\xi_{1}<0.5$}
& \multicolumn{1}{c|}{$0.1<\xi_{2}<0.5$} & \multicolumn{1}{c|}{$0.015<\xi_{3}<0.15$}
\\\hline
$\rm R_{1}$  &1.58\% (1.64\%) &1.09\% &0.46\% \\
\hline	
$\rm R_{2}$  &0.22\% (0.22\%) &0.10\% &0.02\% \\
\hline
$\rm R_{3}$  &13.92\% (13.75\% ) &9.28\% &3.92\% \\
\hline	
\end{tabular}
\end{center}\caption{\label{tab:1d}
The ratios for  $\rm \eta_{c}$ and $\rm \eta_{b}$ (in parentheses ) hadroproductions at LHC for different forward detector acceptances.}
\end{table*}

\subsection{Double diffraction distributions}

In Fig.\ref{fig3:limits}, we exhibit our predictions of  $\rm y^{\eta_{c}}$ and $\rm y^{\eta_{b}}$ distributions for double diffractive hadroproduction in $\rm pp$ collisions at the LHC energies for three different forward detector acceptances. The incident protons which are sources of Pomeron and Reggeon remain undissociated in the final state. The forward and backward detectors are placed at small angles to observe those intact protons while the central detector is located to detect the $\rm \eta_{c}$ or $\rm \eta_{b}$ and other particles. Protons can either emit gluons, Pomeron or Reggeon. When the colliding protons emit gluons, the emerging protons remain dissociated. There is also a case where one proton emits a Reggeon and other proton emits a Pomeron. We can observe that in these differently aforementioned collisions, the  $\rm y^{ \eta_{c}}$ and $\rm y^{ \eta_{b}}$ distributions are symmetric with respect to the mid-rapidity $\rm y^{\eta_{c} }=0$ and $\rm y^{\eta_{b} }=0$ where $\eta_{c}$, $\eta_{b}$ and other unknown particles ($\rm X$ and $\rm X'$) are detected.
This symmetry is due to equal forward and backward rapidities where the colliding protons emits either the same particles such as gluon, Pomeron and Reggeon, or differnt particles (Reggeon from one proton and Pomeron from other proton). The Pomeron and Reggeon behaving like composite particles will emit in their turns diffractive gluons for the hard interactions. The total Pomeron-Reggeon, Reggeon-Pomeron and Reggeon-Reggeon contribution also shows a symmetric distribution. $\rm y^{\eta_{c}}$ and $\rm y^{\eta_{b}}$ distributions for non diffractive process largely predominate over the diffractive processes. The $\rm y^{\eta_{c}}$ distributions from Pomeron-Pomeron (Reggeon-Reggeon) interactions are the lowest one for the forward acceptance detector $0.0015<\xi_{1}<0.5$ and $0.1<\xi_{2}<0.5$ ($0.015<\xi_{3}<0.15$). The Reggeon contribution is sensitive to forward detector acceptances such as $\rm \xi_{1,2}$. It becomes clearly dominant for $0.0015<\xi_{1}<0.5$. The $\rm y^{\eta_{c}}$ and $\rm y^{ \eta_{b}}$ distributions for the double diffractive dissociation have maximums concentrated at midrapidities. The contribution of Reggeon-Reggeon and Reggeon-Pomeron interactions can not be disregarded over the Pomeron-Pomeron interaction in some regions where the $\rm \xi_{1,2}$-cuts are applied. The $\rm y^{\eta_{b}}$ distributions are significant than that of The $\rm y^{\eta_{c}}$ distribution ones for the three forward detector acceptances. By measuring these distributions, we should be able to investigate the Reggeon contribution at the LHC data.
\begin{figure}[htp]
\centering
%\begin{minipage}[t]{4.0cm}
\includegraphics[height=4.8cm,width=4.4cm]{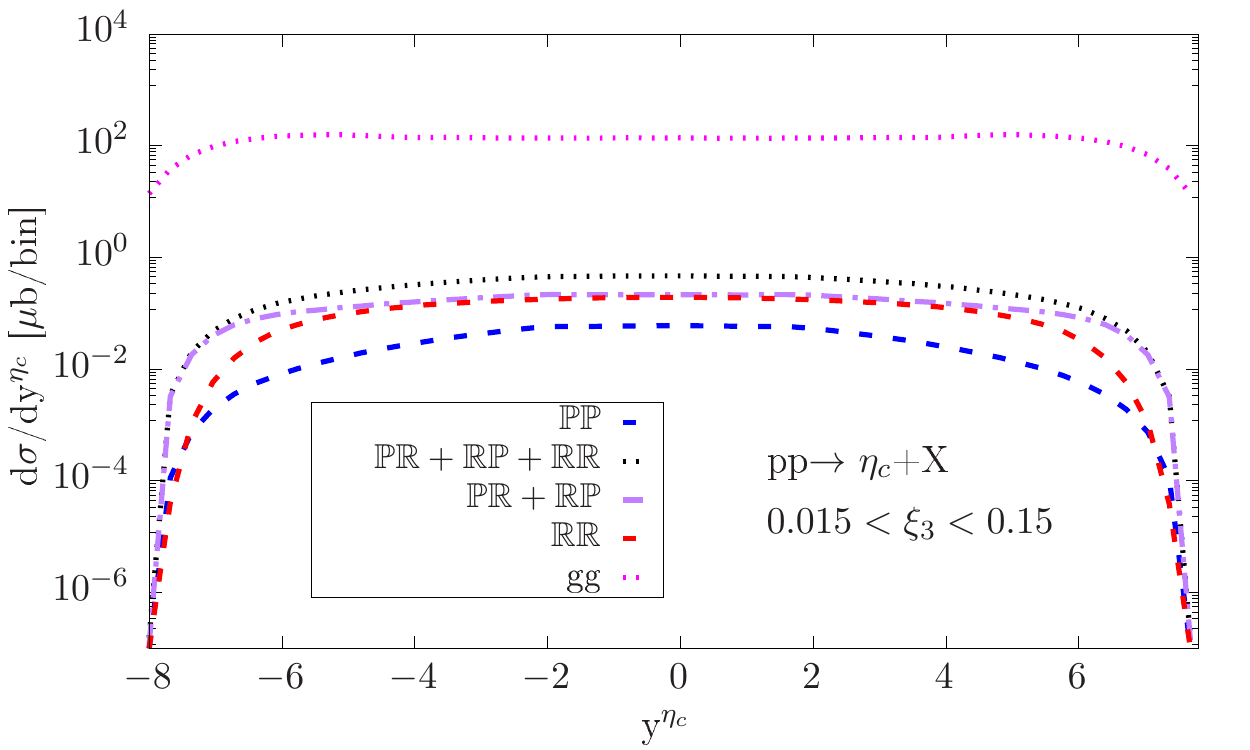}
\includegraphics[height=4.8cm,width=4.4cm]{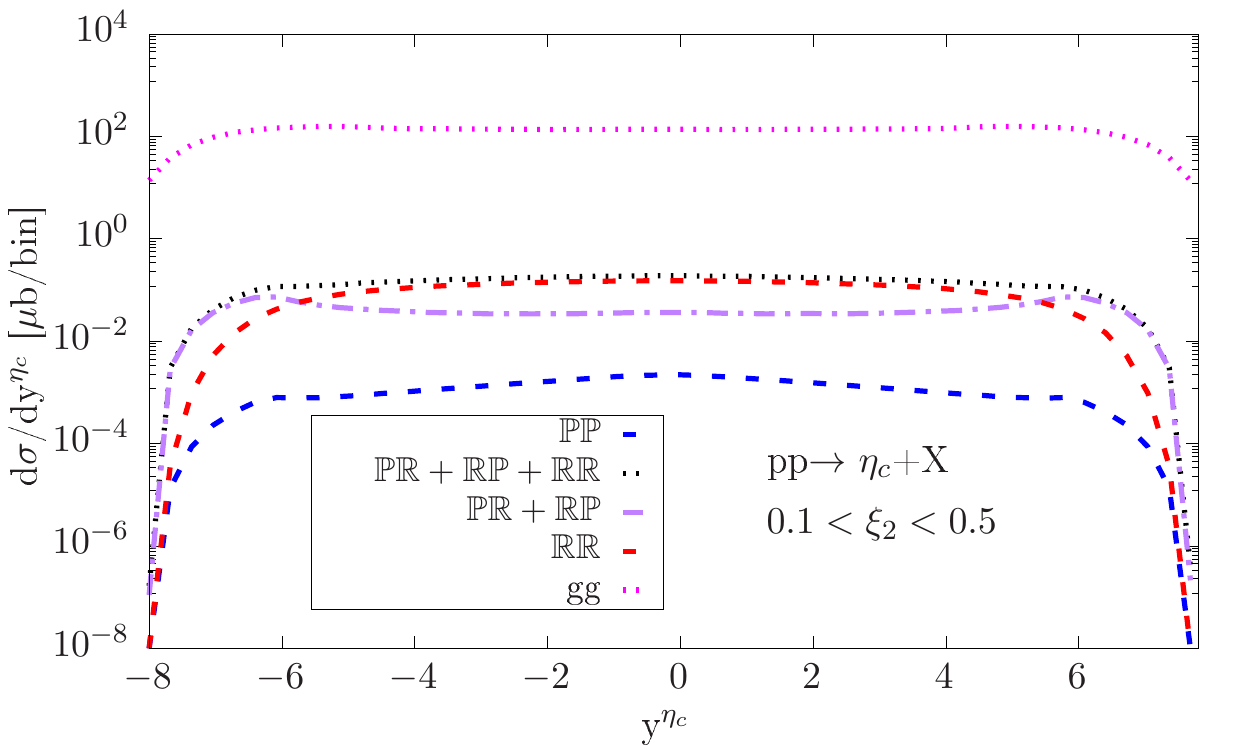}
\includegraphics[height=4.8cm,width=4.4cm]{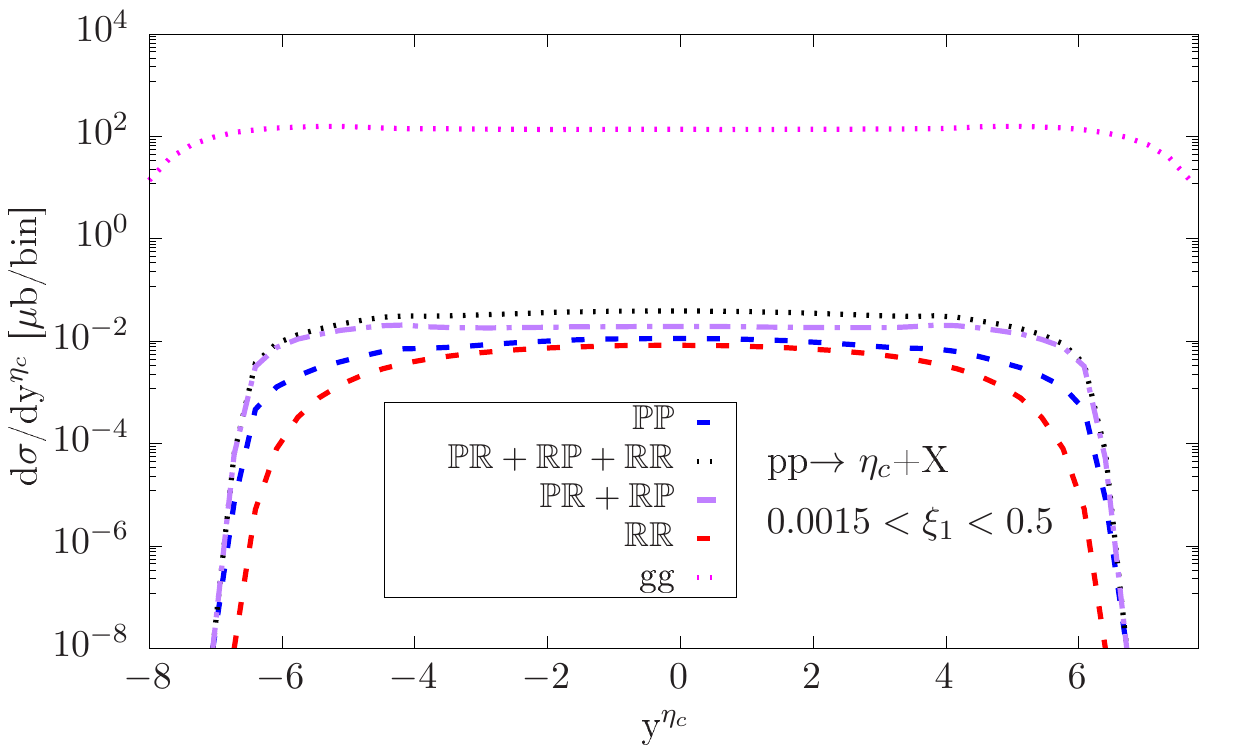}
\includegraphics[height=4.8cm,width=4.4cm]{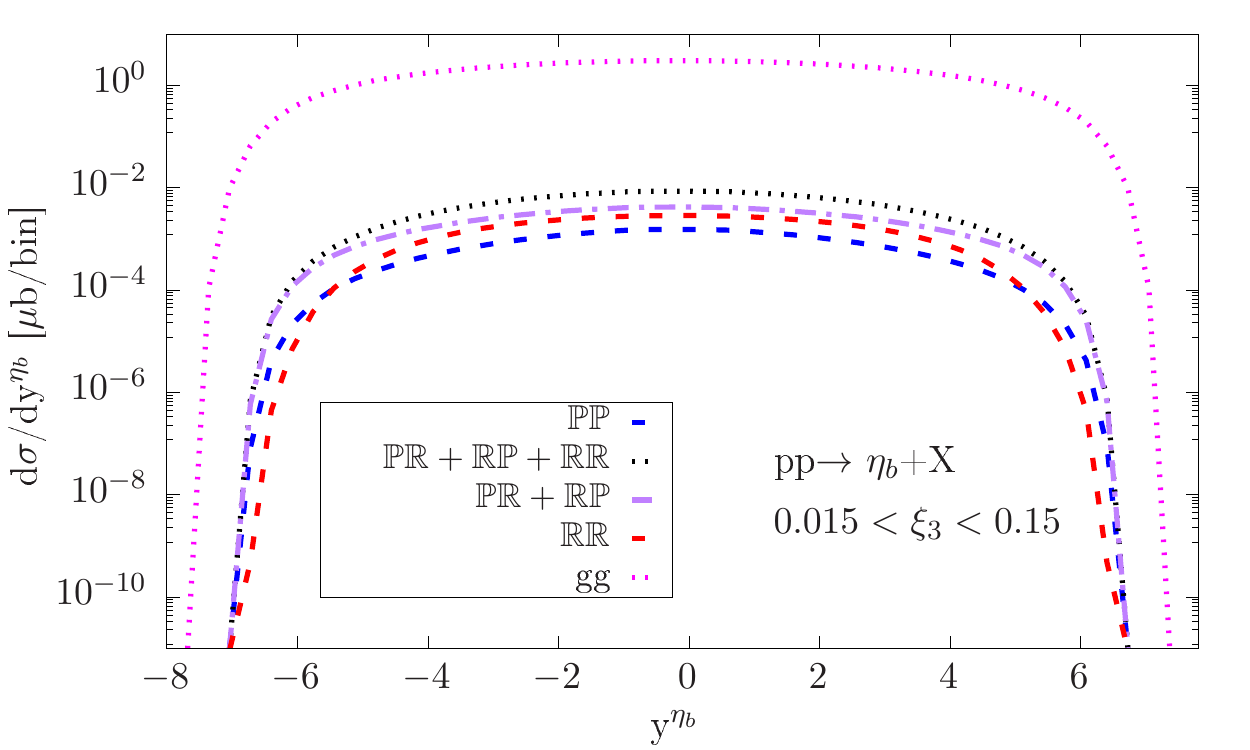}
%\end{minipage}
\caption{\normalsize (color online) The $\rm y^{\eta_{c}}$ and $\rm y^{\eta_{b}}$  distributions for the $\mathbb{PP}$ (blue dashed line), $\mathbb{PP +PR+RP+RR}$ (black dotted line), $\mathbb{ PR+RP}$ (purple dash dotted line), $\mathbb{ RR}$ (red dashed line) and $\rm gg$ (magenta dotted line) in DD processes.}
\label{fig3:limits}
\end{figure}

In Fig.\ref{fig4:limits} , we have plotted $x^{\eta_{c}}_{\mathbb{P}_{1}}$, $x^{\eta_{c}}_{\mathbb{P}_{2}}$, $x^{\eta_{b}}_{\mathbb{P}_{1}}$ and $x^{\eta_{b}}_{\mathbb{P}_{2}}$ distributions for DD processes. We have noticed the distributions of two colliding protons in DD process are slightly similar because their proton momentum loss are closely equal. The Reggeon-Reggeon contribution increases for low range of $x_{\mathbb{P}_{1}}$ and $x_{\mathbb{P}_{2}}$. It becomes flat for large $x_{\mathbb{P}_{1}}$ and $x_{\mathbb{P}_{2}}$ ranges where its contribution is non-negligible. The Pomeron-Pomeron contribution continuously decreases for low and large ranges of $x_{\mathbb{P}_{1}}$ and $x_{\mathbb{P}_{2}}$, meanwhile the Pomeron-Reggeon contribution decreases for low range and turns out to be flat for large range. Reggeon-Reggeon contribution is useful at large $x_{\mathbb{P}_{1}}$ and $x_{\mathbb{P}_{2}}$ over the Pomeron-Pomeron contribution one and can not be neglected. Reggeon contributions dominate for large values of proton momentum loss while the Pomeron exchange is still dominant for small values \cite{Marquet:2016ulz}. The Reggeon contribution sensitivity can be increased near the edge of the proton forward detector acceptance and it becomes evidently dominant.
\begin{figure}[htp]
\centering
%\begin{minipage}[t]{4.0cm}
\includegraphics[height=4.8cm,width=4.4cm]{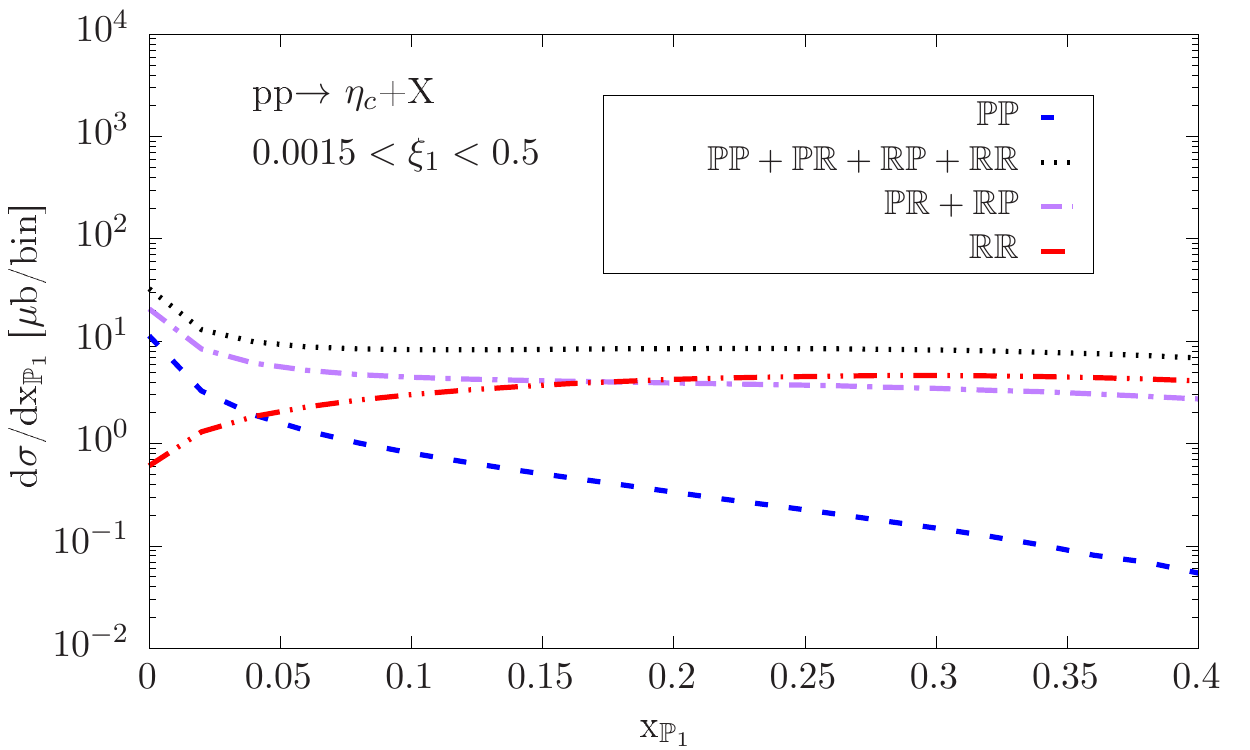}
\includegraphics[height=4.8cm,width=4.4cm]{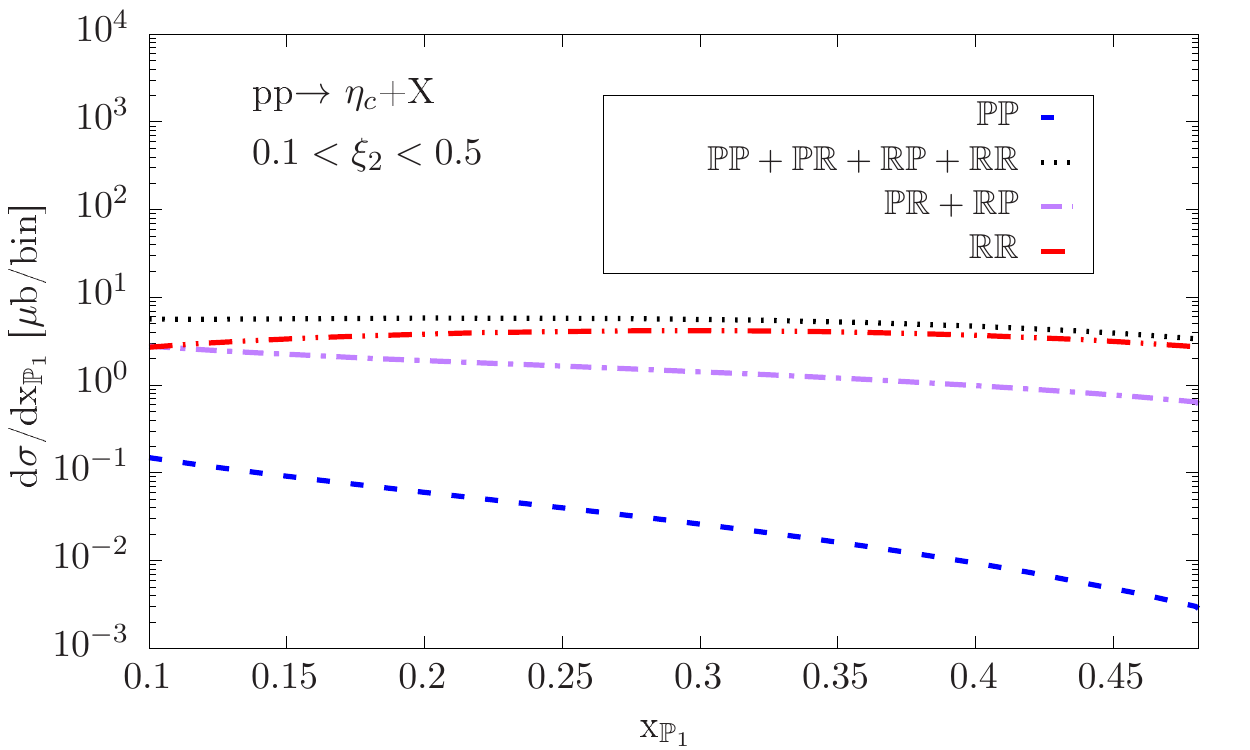}
\includegraphics[height=4.8cm,width=4.4cm]{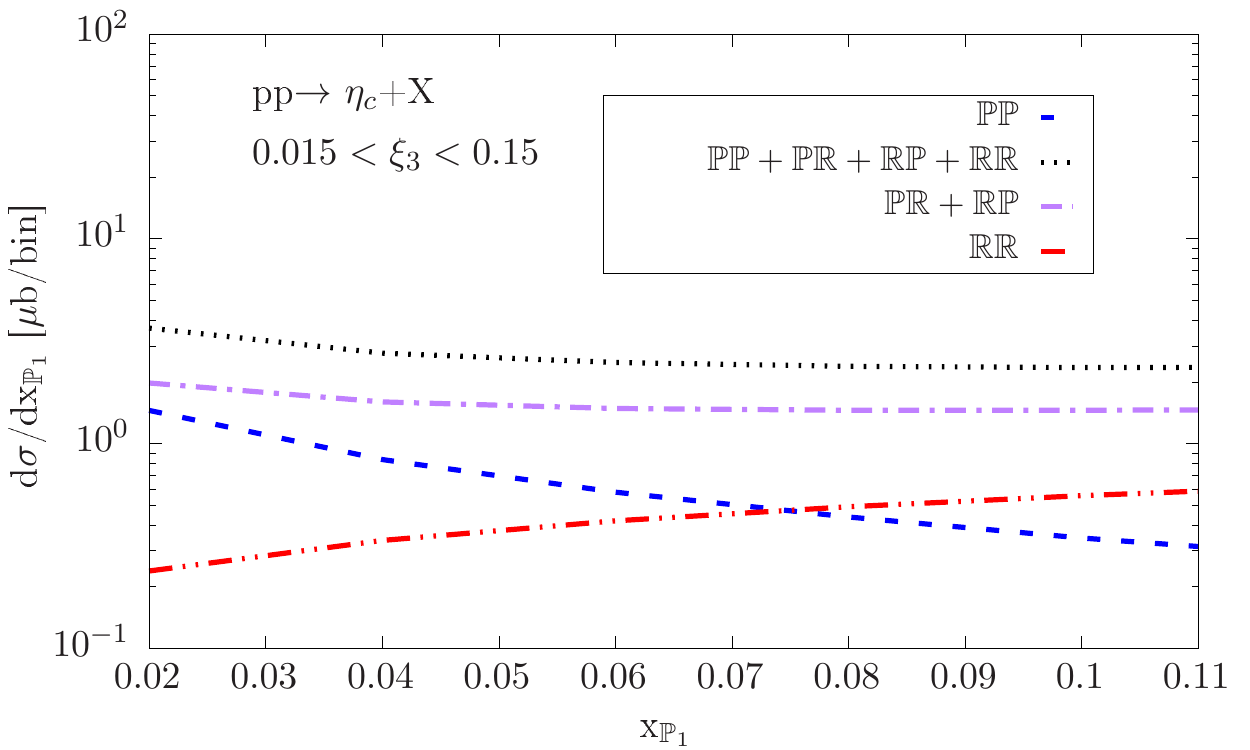}
\includegraphics[height=4.8cm,width=4.4cm]{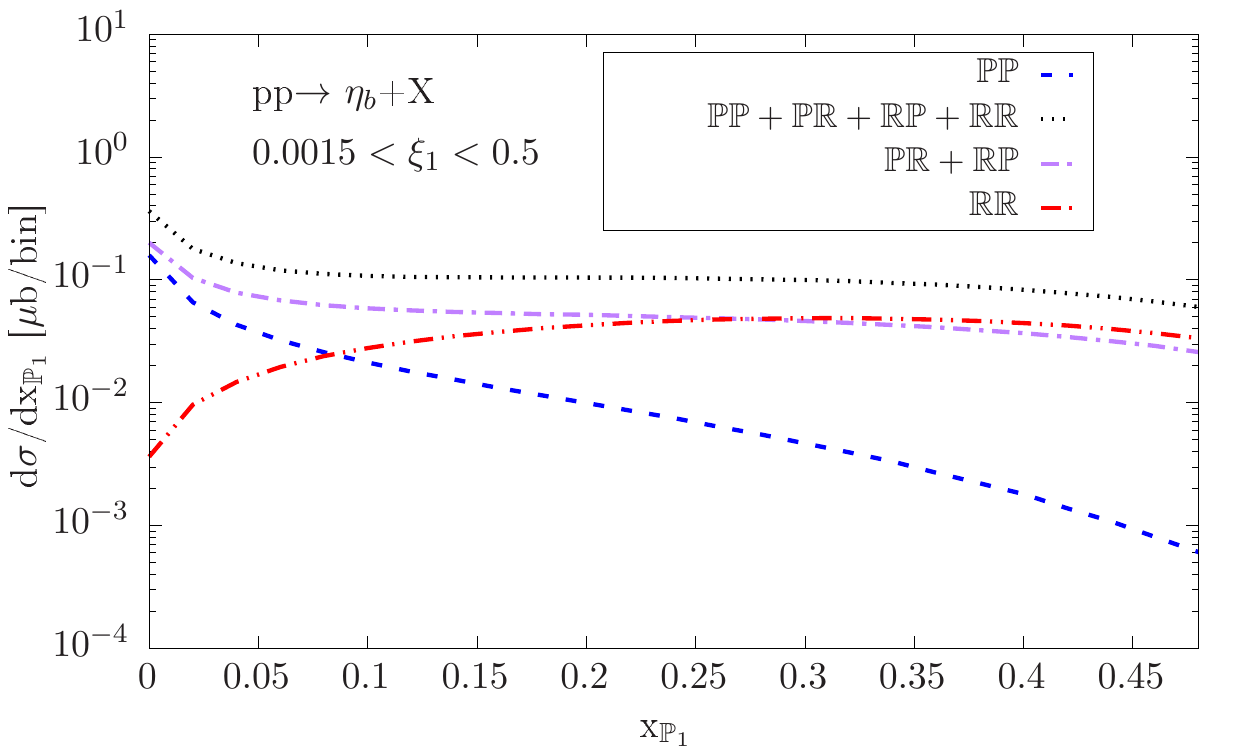}\\
\includegraphics[height=4.8cm,width=4.4cm]{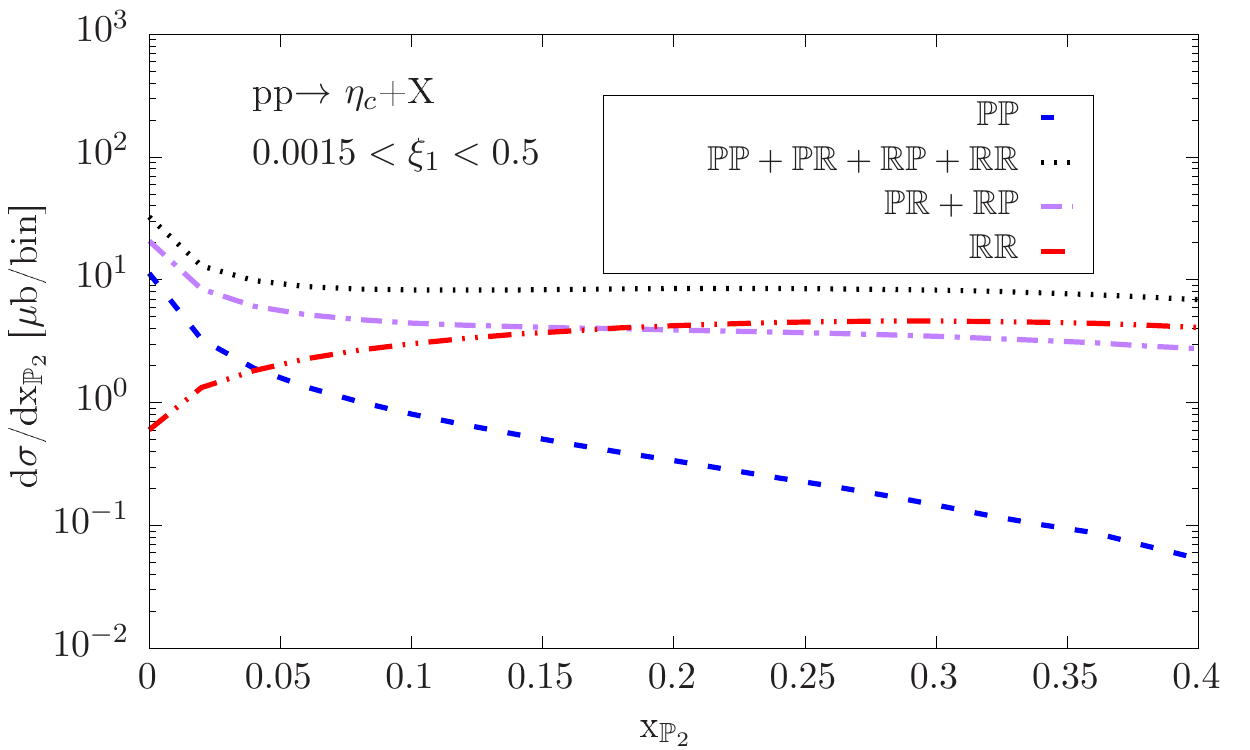}
\includegraphics[height=4.8cm,width=4.4cm]{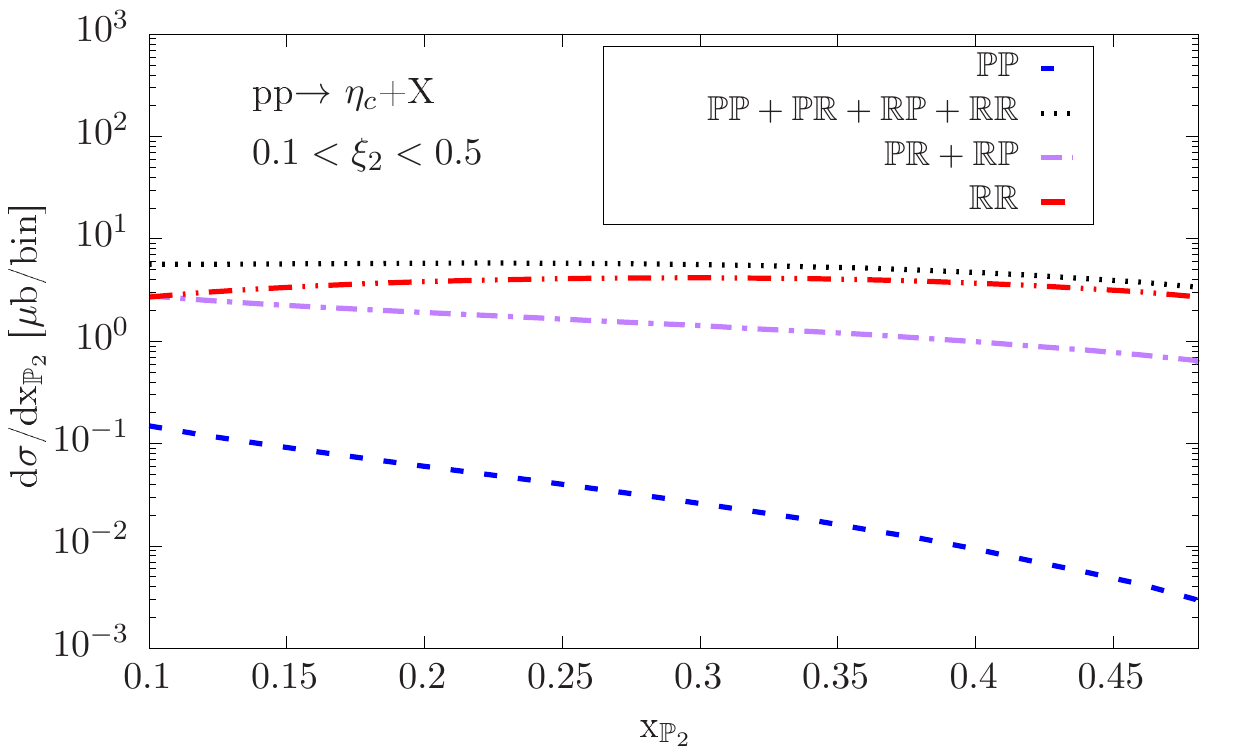}
\includegraphics[height=4.8cm,width=4.4cm]{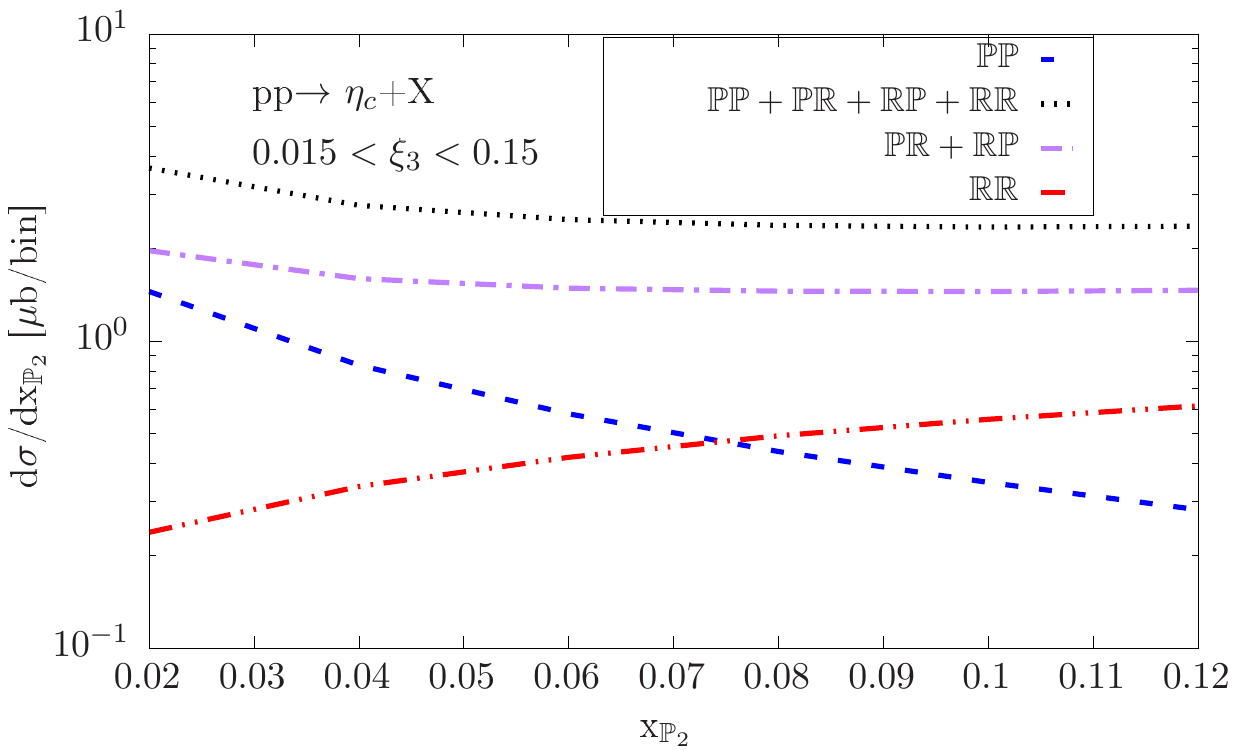}
\includegraphics[height=4.8cm,width=4.4cm]{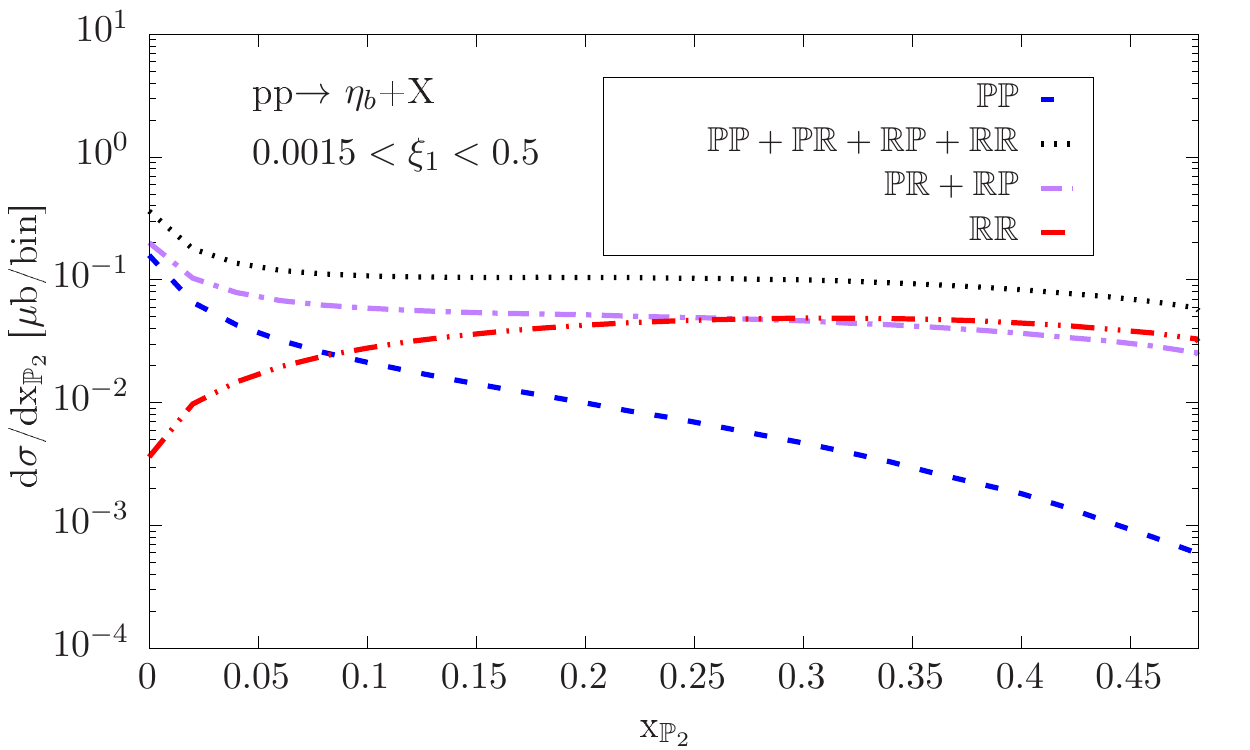}
%\end{minipage}
\caption{ \normalsize (color online)
The $x^{\eta_{c}}_{\mathbb{P}_{1}}$ and $x^{\eta_{b}}_{\mathbb{P}_{2}}$ distributions (top panel) and, $x^{\eta_{c}}_{\mathbb{P}_{2}}$ and $x^{\eta_{b}}_{\mathbb{P}_{1}}$ distributions ( bottom panel) for the  $\mathbb{PP}$ (blue dashed line), $\mathbb{PP +PR+RP+RR}$ (black dotted line), $\mathbb{ PR+RP}$ (purple dash dotted line) and $\mathbb{ RR}$ (red dashed line) in DD processes.}
\label{fig4:limits}
\end{figure}

In Fig.\ref{fig5:limits}, we have presented the $\beta_{1}$ and $\beta_{2}$ distributions of $\eta_{c}$ for three different forward detector acceptances ($\rm \xi_{1,2,3}$). We have also plotted for one forward detector acceptance ($\rm \xi_{3}$) for $\eta_{b}$. We realize that all the contributions decrease, become flat and decrease again. Except for the Reggeon-Reggeon contribution where the decrease is spread over all range of $\beta_{1}$ and $\beta_{2}$. The Reggeon-Reggeon contribution is dominant for small $\beta_{1}$ and $\beta_{2}$ over the Pomeron-Pomeron contribution. They become comparable for very small $\beta_{1}$ and $\beta_{2}$. However, we also notice that the Pomeron-Pomeron contribution surpasses that of Reggeon-Reggeon one for large $\beta_{1}$ and $\beta_{2}$. For $0.015<\xi_{3}<0.15$ ($0.0015<\xi_{1}<0.5$) the Reggeon-Reggeon  contribution is dominated by the Pomeron-Pomeron one for $\eta_{c}$ ($\eta_{b}$) for small $\beta_{1}$ and $\beta_{2}$. When $\beta_{1}$ and $\beta_{2}$ tend to very small values, the Reggeon-Reggeon contribution is comparable to Pomeron-Pomeron contribution. The behavior of these plots is related to  $ \beta_{1}=\frac{x_{1}}{x_{\mathbb{P}_{1}}}$ and  $ \beta_{2}=\frac{x_{2}}{x_{\mathbb{P}_{2}}}$.
\begin{figure}[htp]
\centering
%\begin{minipage}[t]{4.0cm}
\includegraphics[height=4.8cm,width=4.4cm]{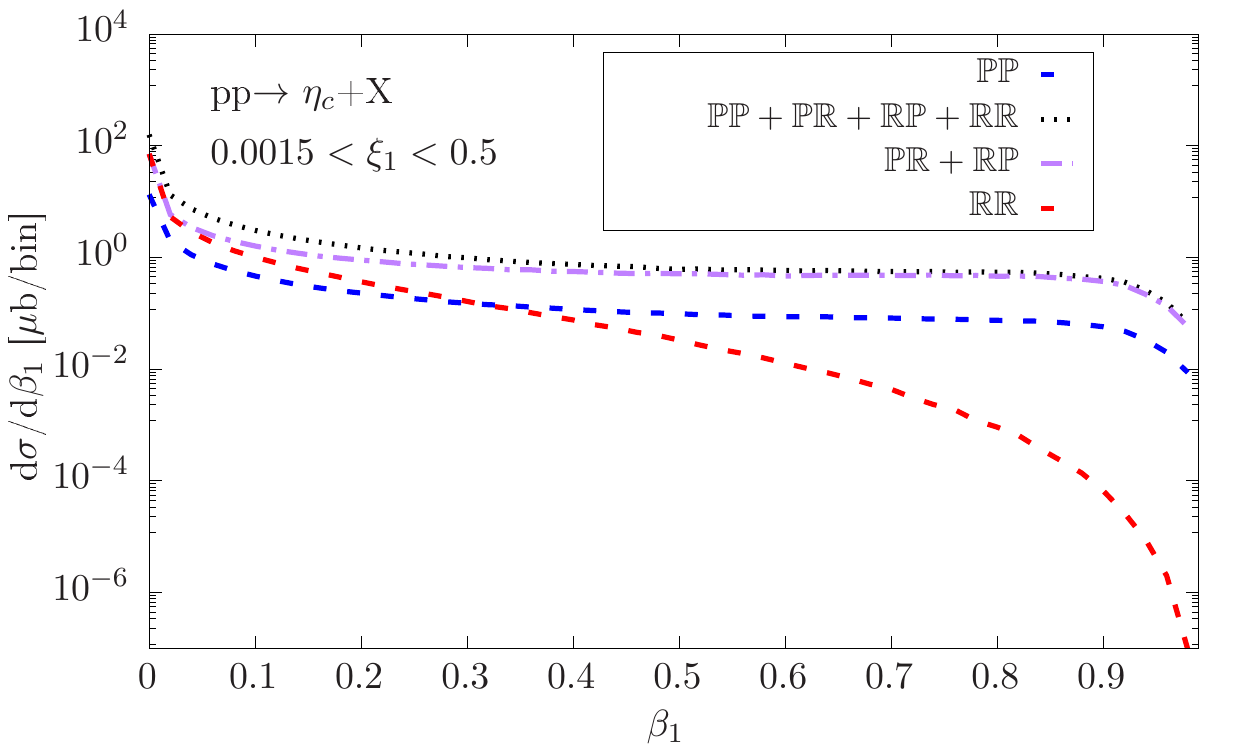}
\includegraphics[height=4.8cm,width=4.4cm]{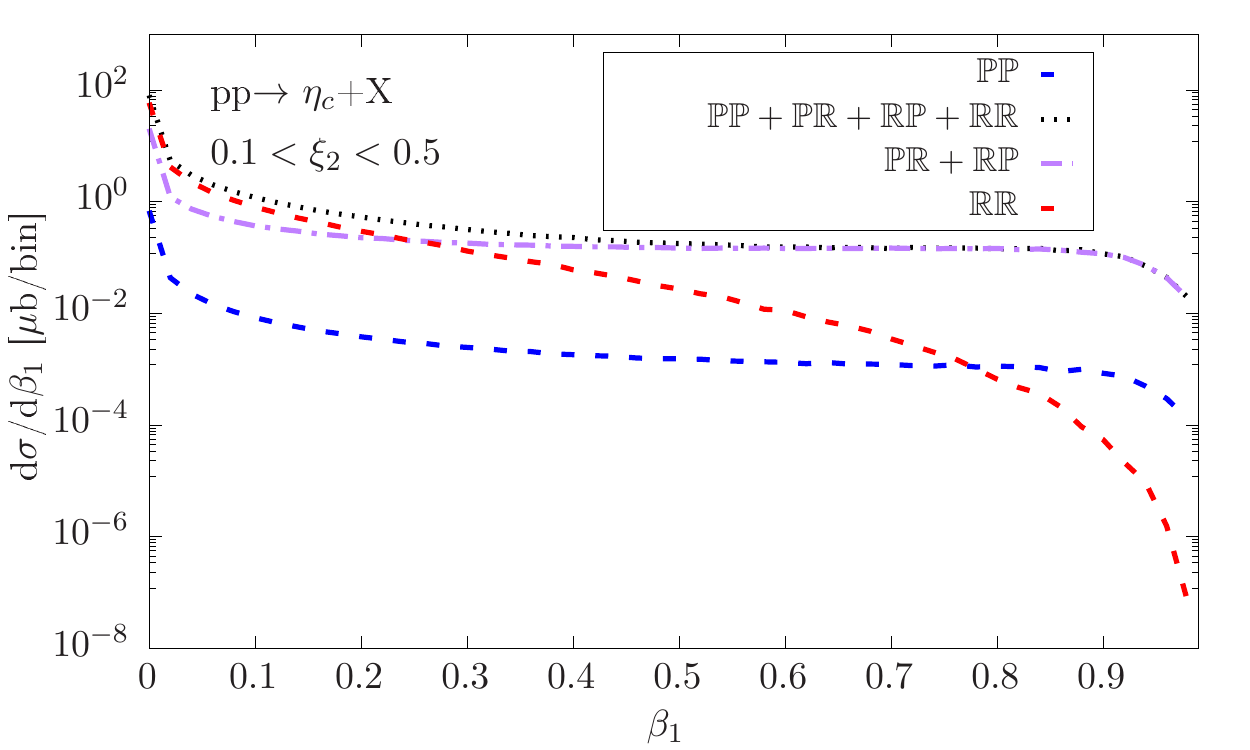}
\includegraphics[height=4.8cm,width=4.4cm]{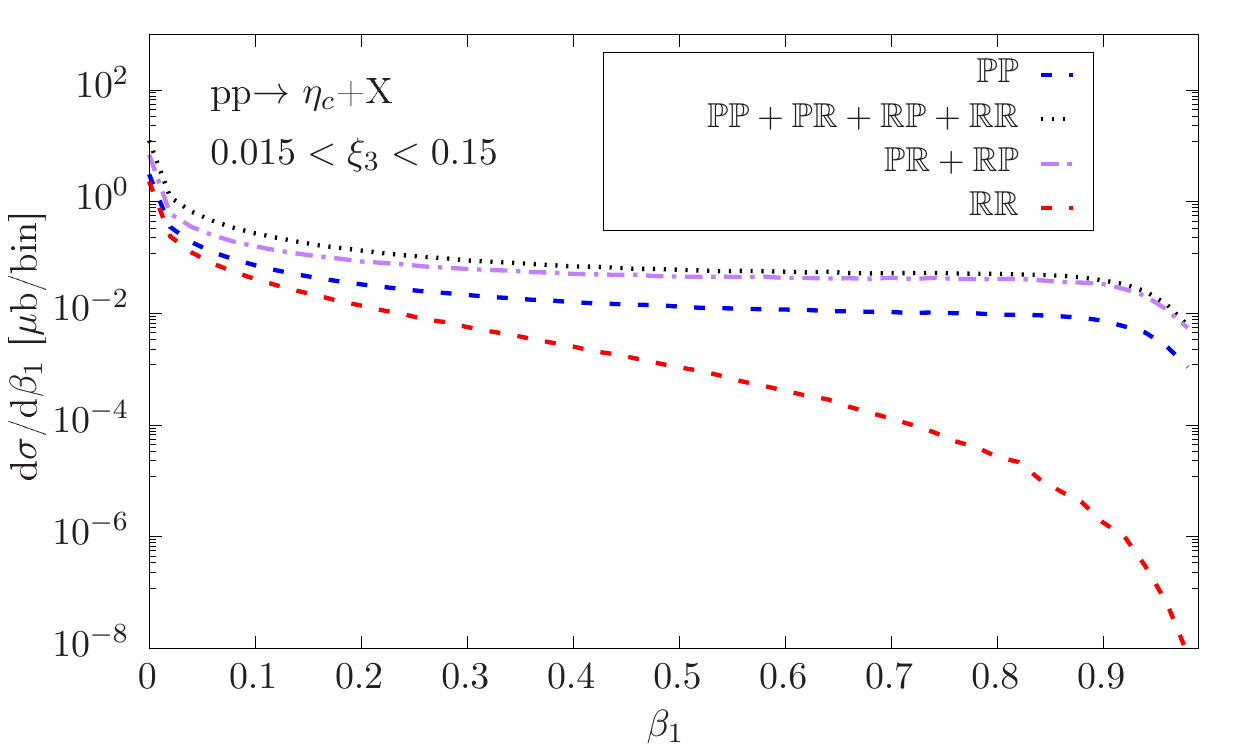}
\includegraphics[height=4.8cm,width=4.4cm]{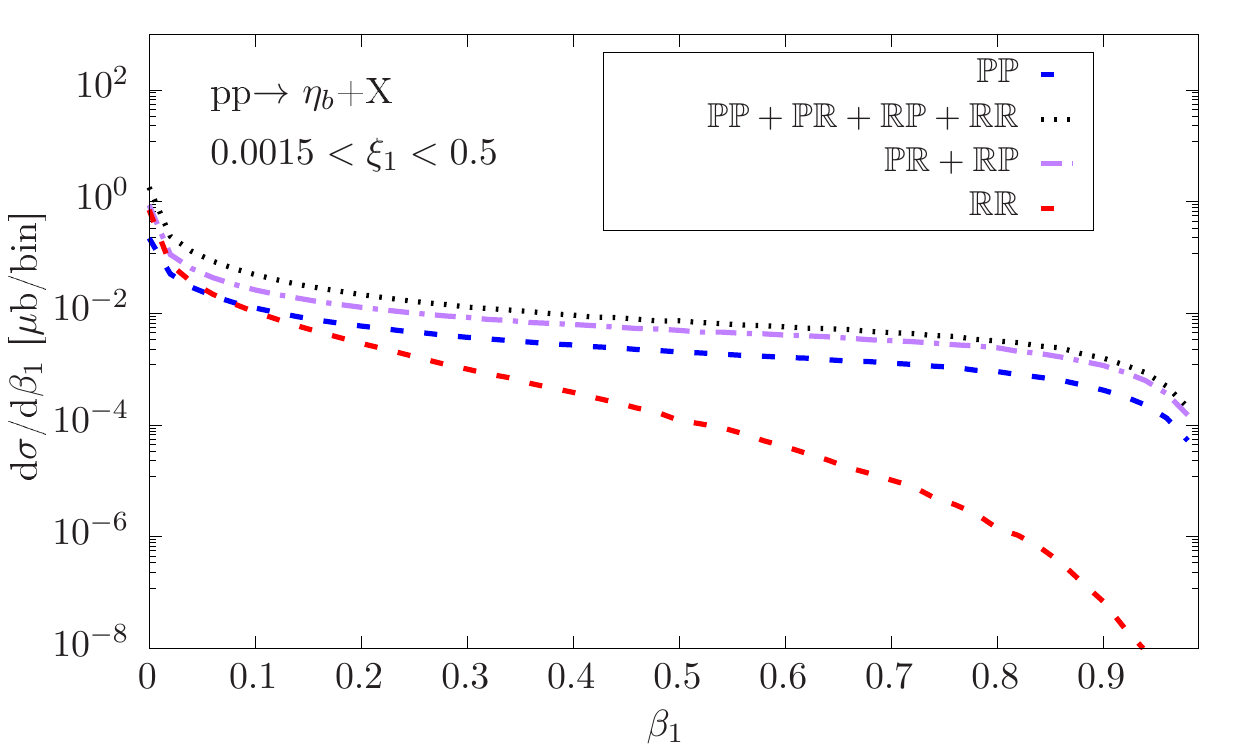}
\includegraphics[height=4.8cm,width=4.4cm]{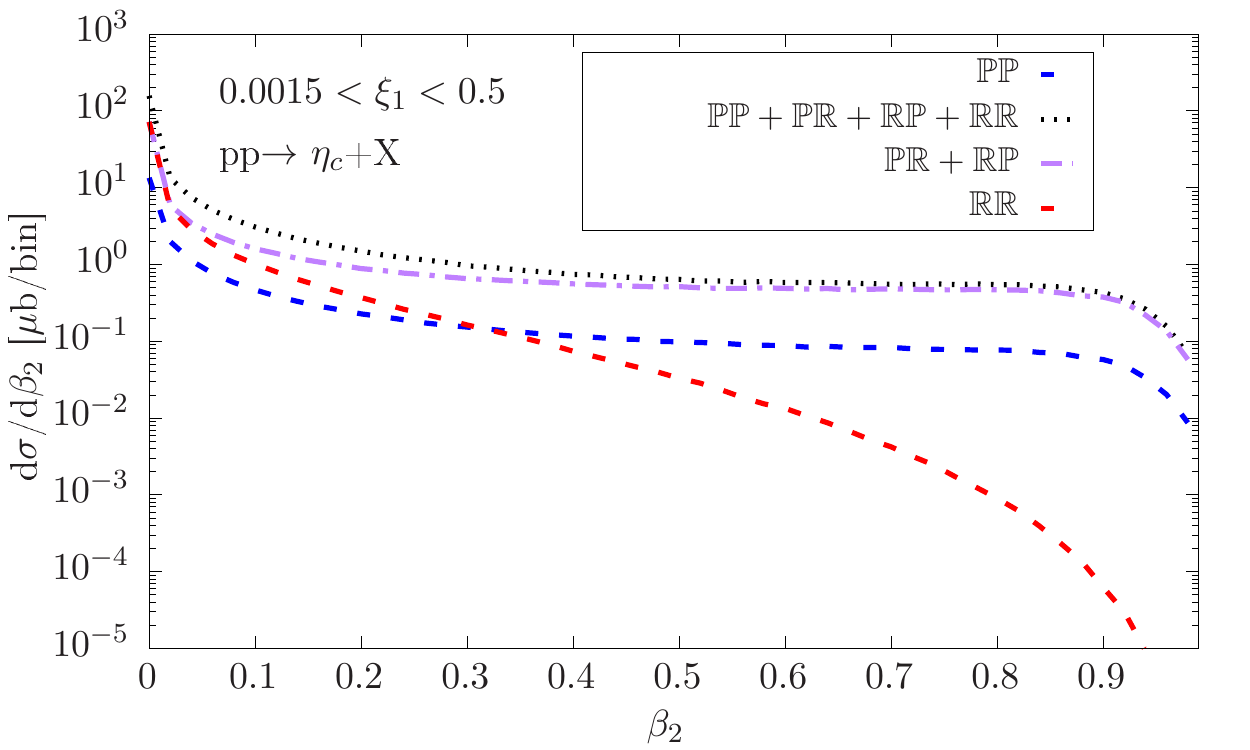}
\includegraphics[height=4.8cm,width=4.4cm]{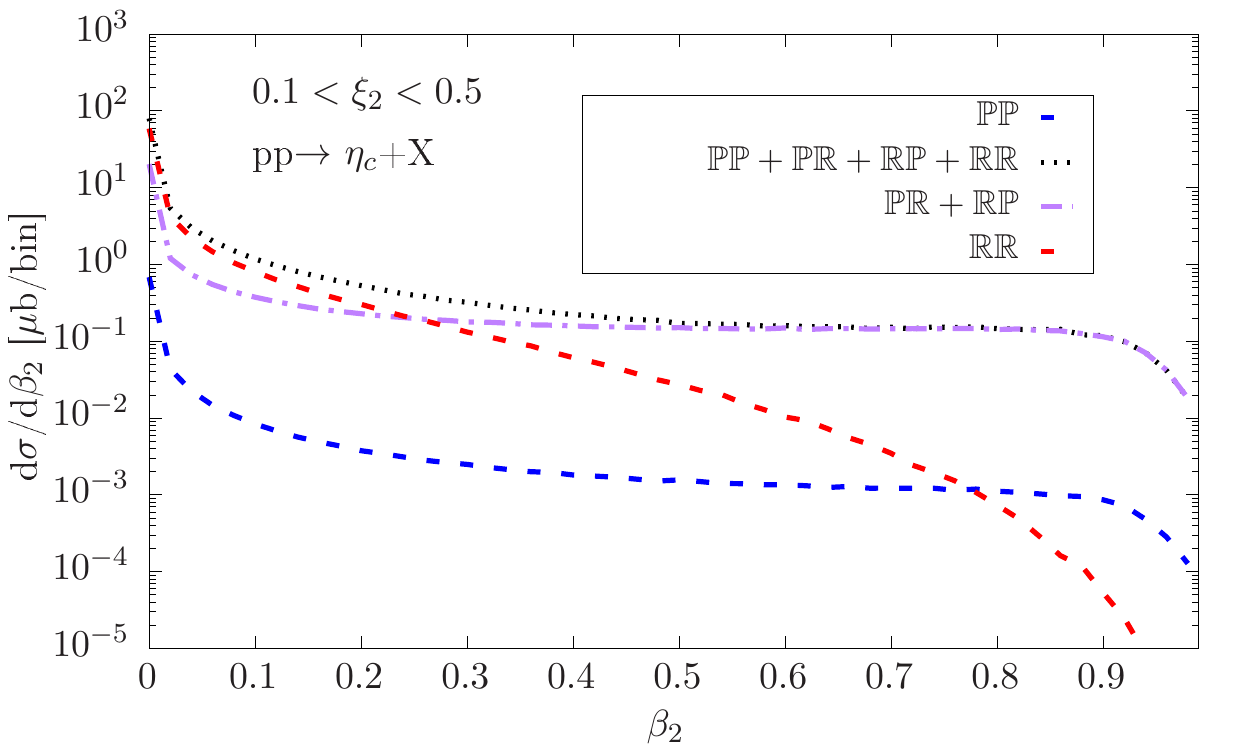}
\includegraphics[height=4.8cm,width=4.4cm]{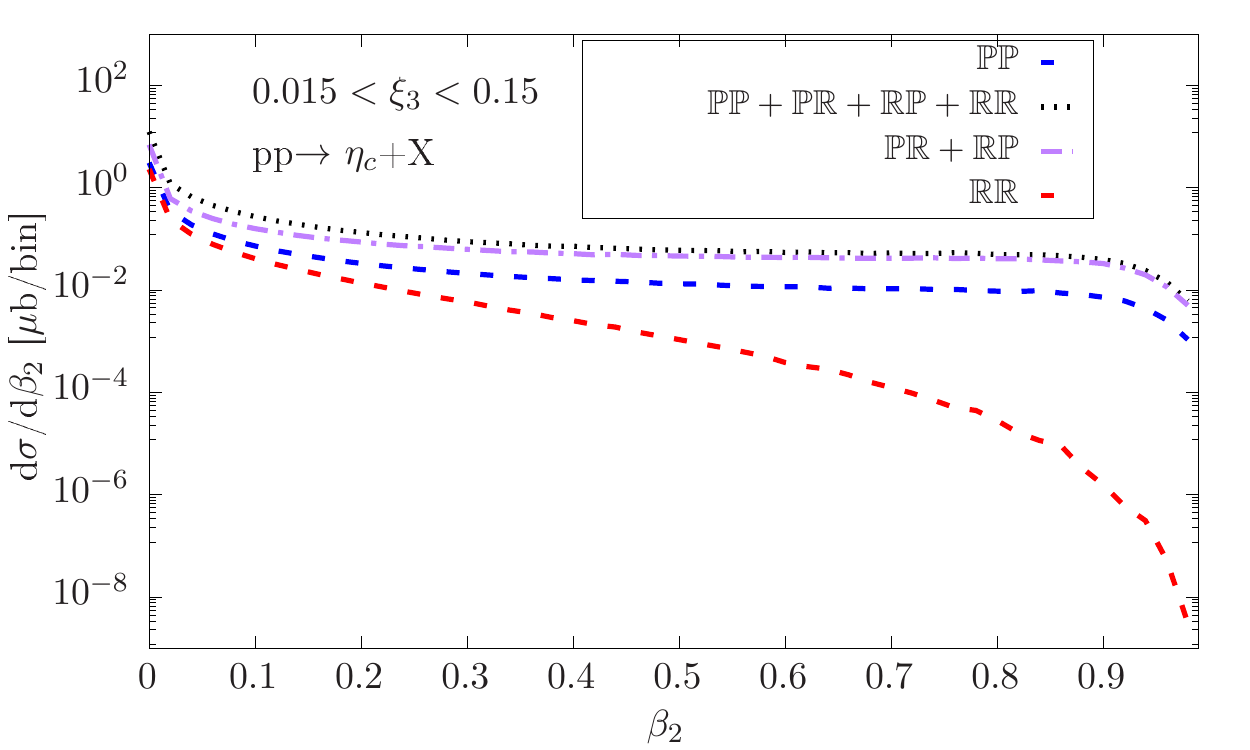}
\includegraphics[height=4.8cm,width=4.4cm]{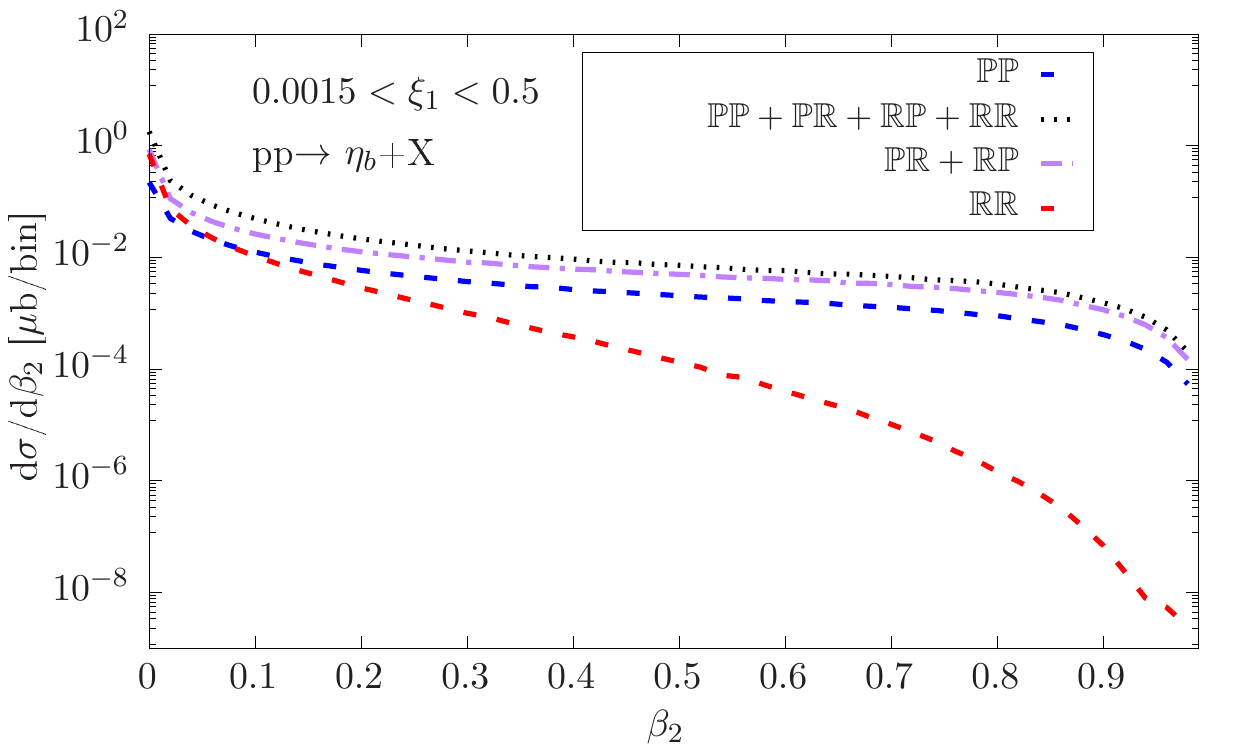}
%\end{minipage}
\caption{ \normalsize (color online)
The $\beta_{1}$ distributions (top panel) and $\beta_{2}$ distributions (bottom panel)  for the $\mathbb{PP}$ (blue dashed line), $\mathbb{PP +PR+RP+RR}$ (black dotted line), $\mathbb{ PR+RP}$ (purple dash dotted line) and $\mathbb{ RR}$ (red dashed line) in DD processes.}
\label{fig5:limits}
\end{figure}

\subsection{Single Diffraction distributions}

In Fig.\ref{fig6:limits}, we have the exhibition of the $\rm y^{\eta_{c}}$  distributions in SD dissociation for three different forward detector acceptances. In this case, one of the two protons emits a gluon and the second proton emits Pomeron or Reggeon with a small squared momentum transfer. What is more, Pomeron or Reggeon can emit also a diffractive gluon before hard scattering. Afterward, gluon and diffractive gluon go into hard collision. The proton emitting the Pomeron or Reggeon remains intact by turning into excited one and is detected by the forward and backward detectors while the proton emitting gluons only dissociates into new system called remnant ($\rm X, X'$) observed by the central detectors. We can see that the  $\rm y^{\eta_{c}}$ distributions are asymmetric with respect to the mid-rapidity  $\rm y^{\eta_{c}}=0$ for $\mathbb{P}$p and $\mathbb{R}$p contributions. This asymmetry is caused by the inequality in forward and backward rapidities. $\rm y^{ \eta_{c} }$  distributions for non diffractive process, where the two protons emit gluons only, largely dominate over the the diffractive processes. Its distibution is symmetric. The $\rm y^{\eta_{c}}$ distributions from Reggeon  interactions dominate over the Pomeron ones for small values of rapidities. Nevertheless, Pomeron contribution is important for  large values of rapidities. The $\rm y^{\eta_{b}}$ distribution is displayed for $0.0015<\xi_{1}<0.5$. Its distribution is more important than that of $\rm y^{ \eta_{c} }$. The $\rm y^{ \eta_{c}}$ and $\rm y^{\eta_{b}}$ distributions for the single diffractive dissociation have maximums shifted to forward and backward rapidities with respect to the non-diffractive case. The constraint on Reggeon distribution at LHC should enhance the theoretical predictions for $\rm \eta_{c}$ and $\rm \eta_{b}$.
\begin{figure}[htp]
\centering
%\begin{minipage}[t]{4.0cm}
\includegraphics[height=4.8cm,width=4.4cm]{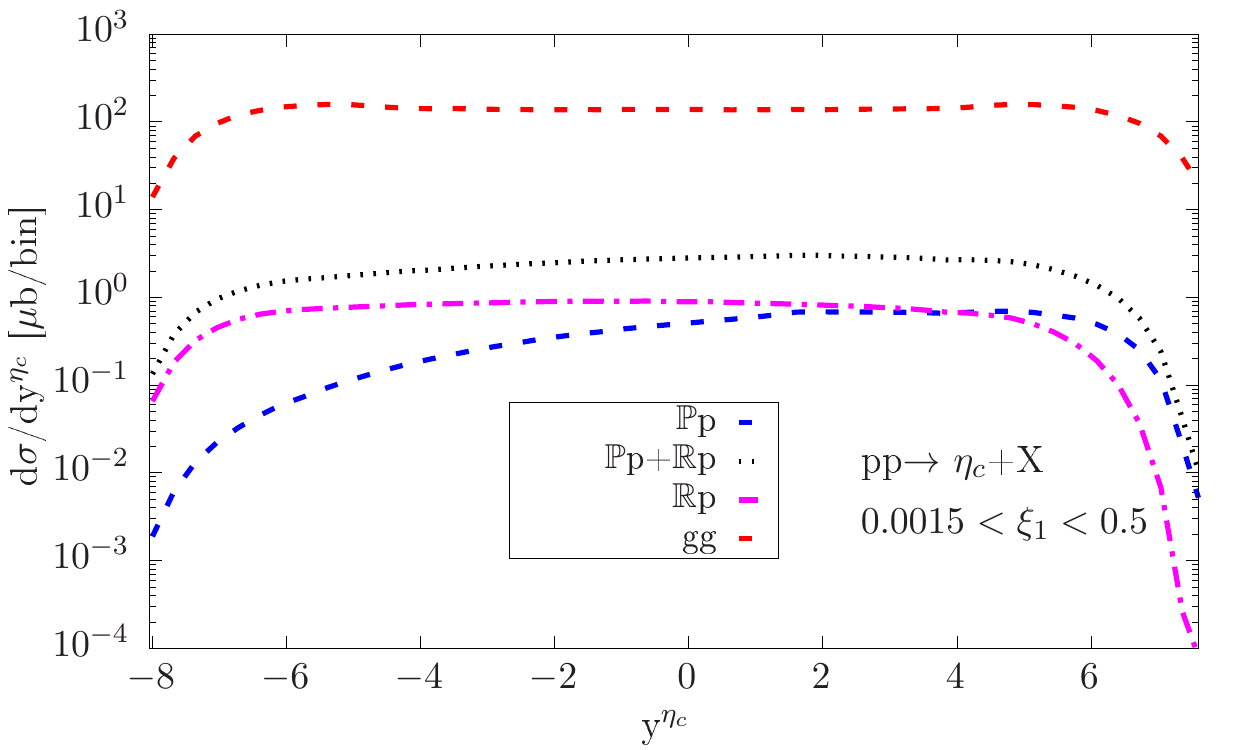}
\includegraphics[height=4.8cm,width=4.4cm]{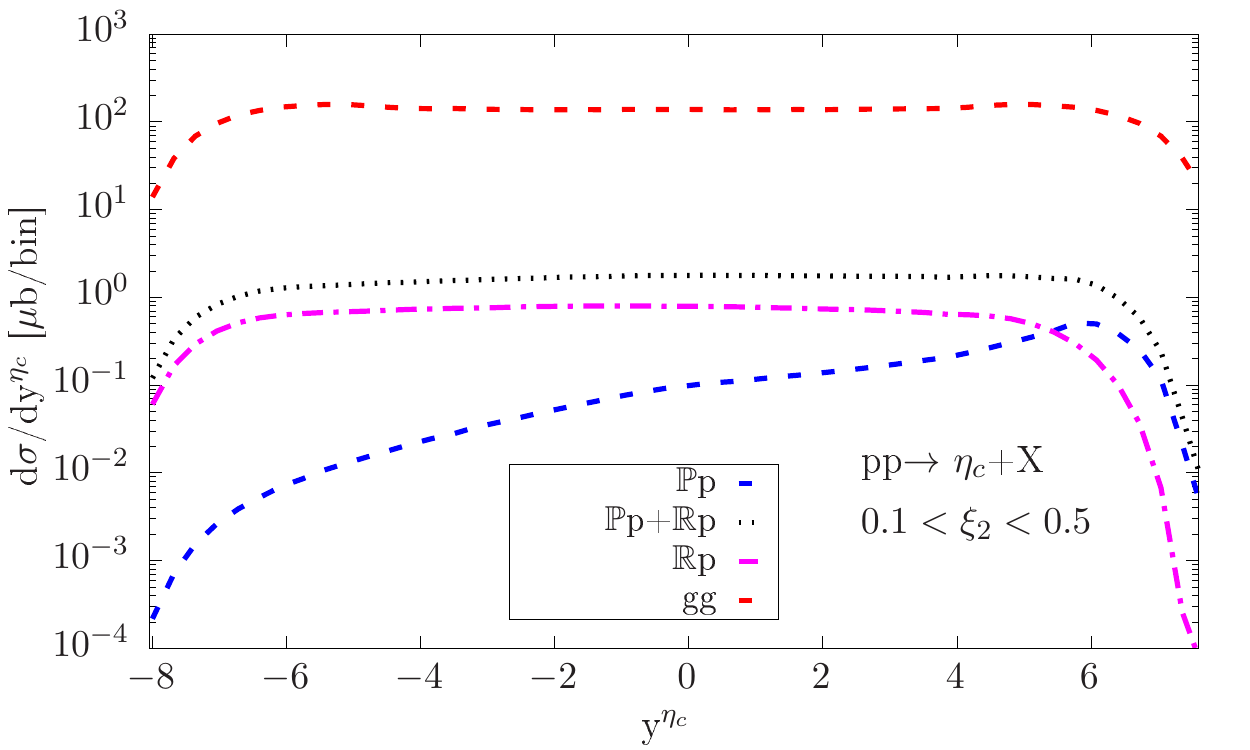}
\includegraphics[height=4.8cm,width=4.4cm]{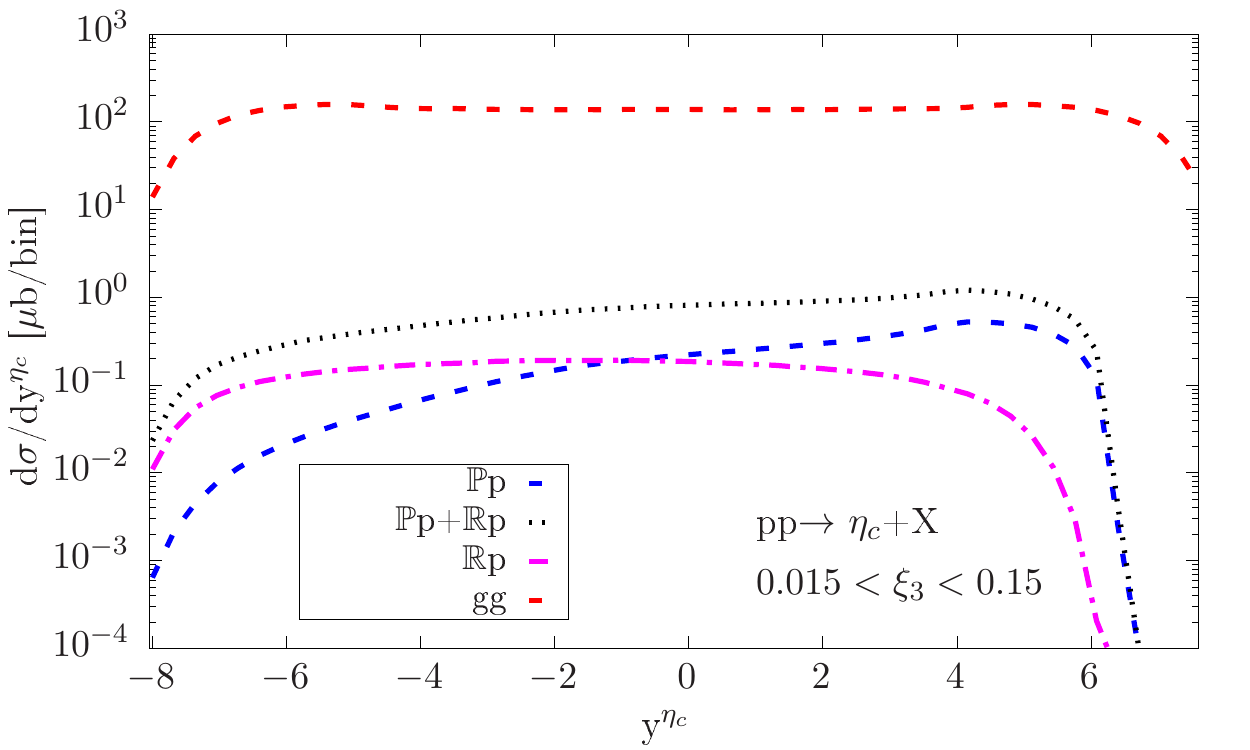}
\includegraphics[height=4.8cm,width=4.4cm]{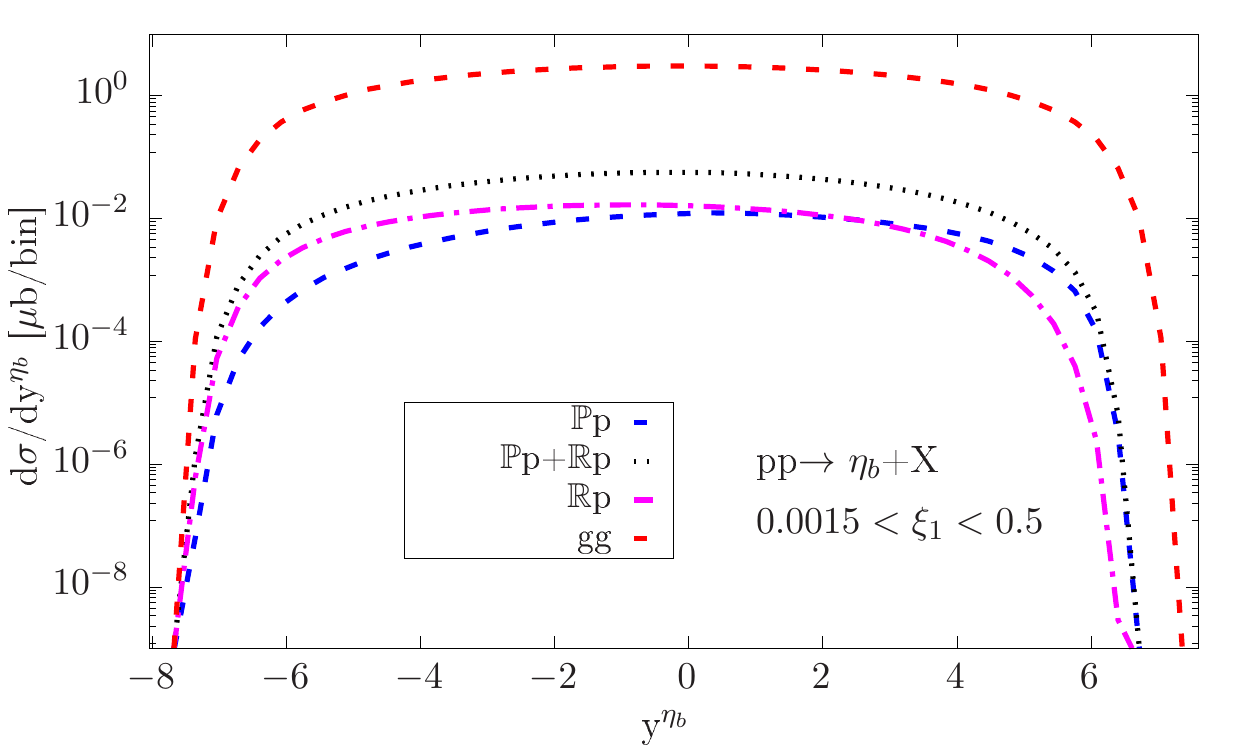}
%\end{minipage}
\caption{ \normalsize (color online)
The $\rm y^{\eta_{c}}$ and $\rm y^{\eta_{b}}$ distributions for the $\mathbb{P}$p (blue dashed line), $\mathbb{P}$p + $\mathbb{R}$p (black dotted line), $\mathbb{ R}$p (purple dash dotted line) and $\rm gg$ (red dotted line) in SD processes.}
\label{fig6:limits}
\end{figure}

In Fig.\ref{fig7:limits} , we have plotted the $x^{\eta_{c}}_{\mathbb{P}}$ and $x^{\eta_{b}}_{\mathbb{P}}$ distributions for SD processes. The $\mathbb{R}$p  contribution increases for low range of $x_{\mathbb{P}}$. It becomes slightly flat for large range where its contribution is non-negligible. The $\mathbb{P}$p  contribution always  decreases for low and large ranges of $x_{\mathbb{P}}$. The $\mathbb{R}$p contributions dominate for large values of proton momentum loss while the  $\mathbb{P}$p is still dominant at small values \cite{Marquet:2016ulz}. The Reggeon contribution sensitivity can be increased near the edge of the proton forward detector acceptance and it becomes evidently dominant.
\begin{figure}[htp]
\centering
%\begin{minipage}[t]{4.0cm}
\includegraphics[height=4.8cm,width=4.4cm]{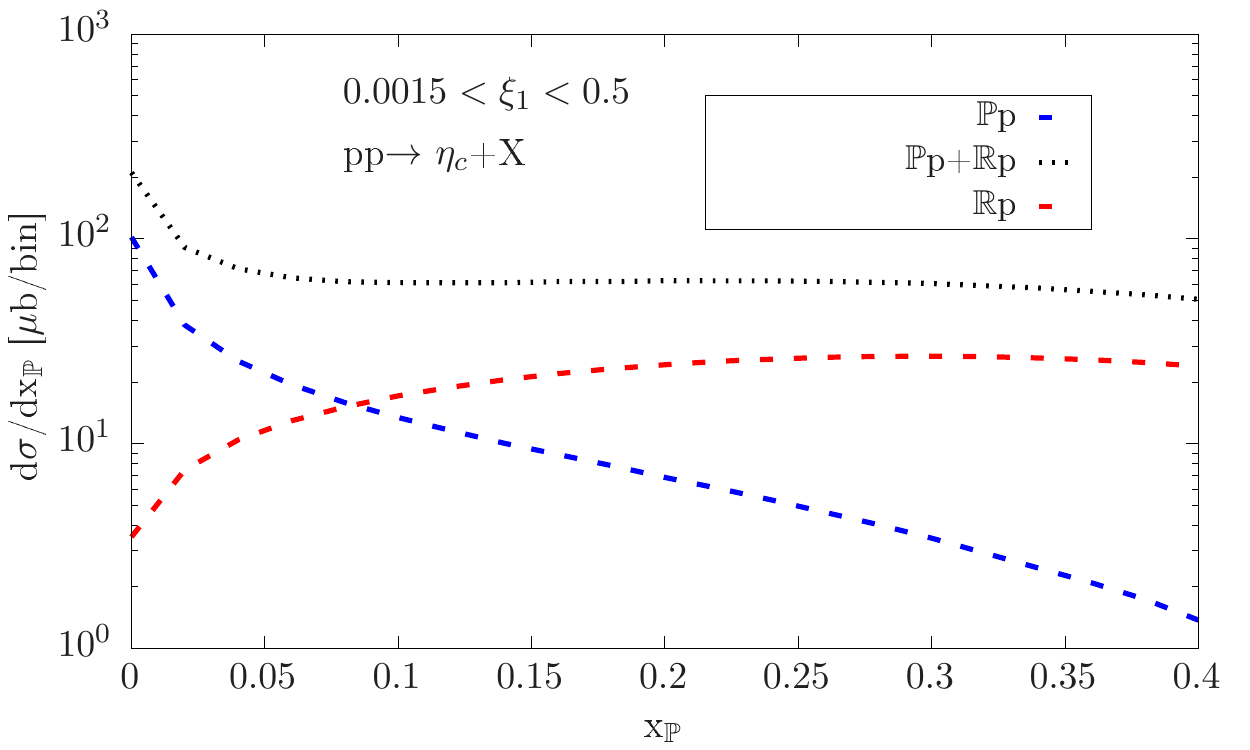}
\includegraphics[height=4.8cm,width=4.4cm]{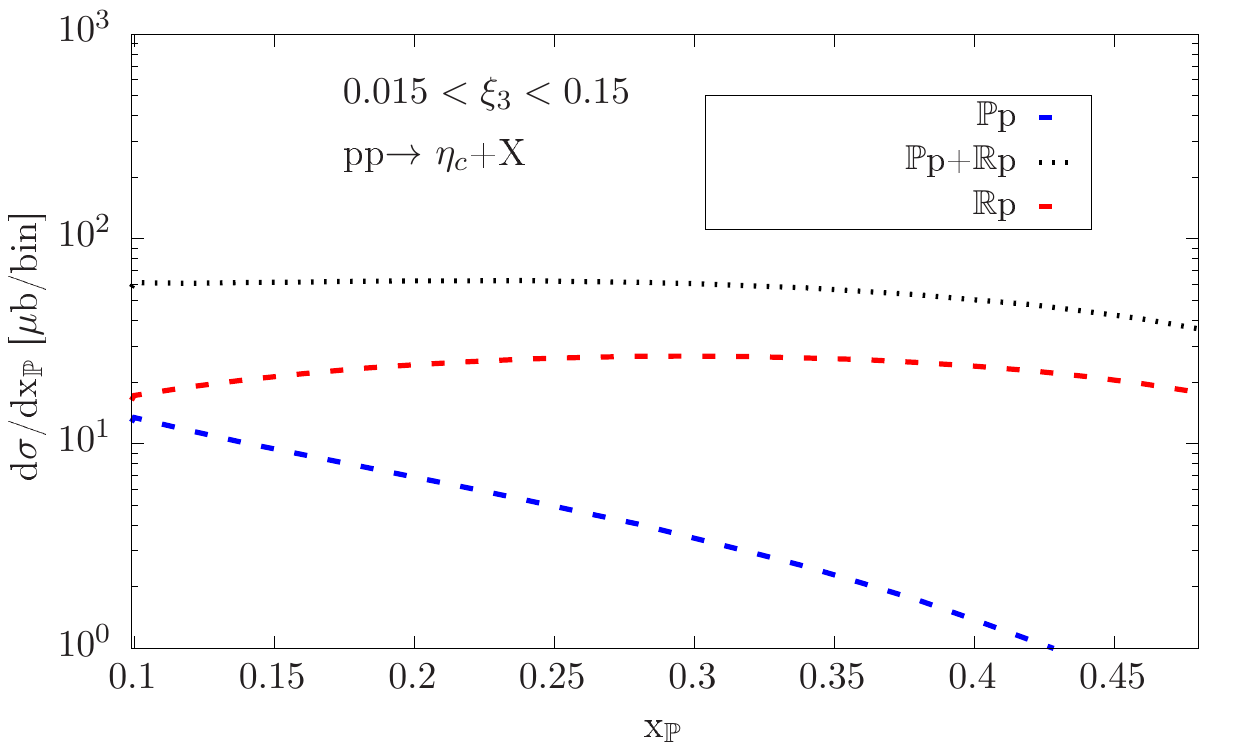}
\includegraphics[height=4.8cm,width=4.4cm]{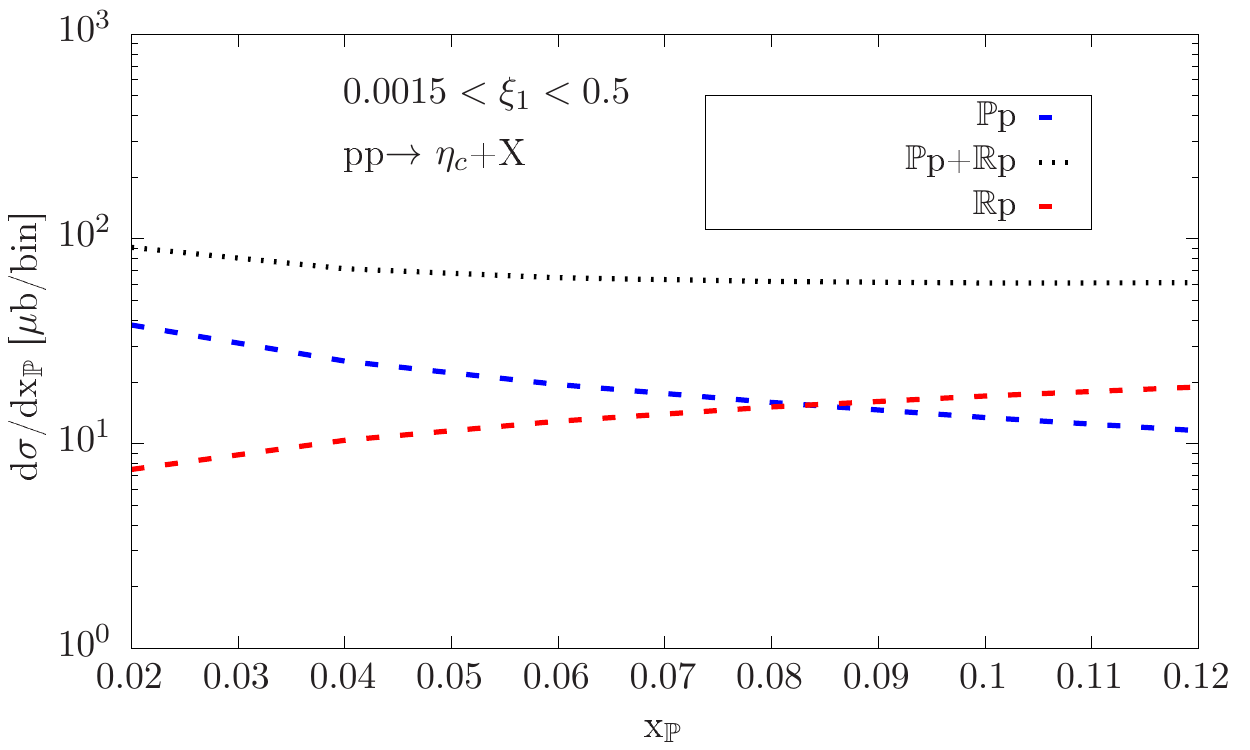}
\includegraphics[height=4.8cm,width=4.4cm]{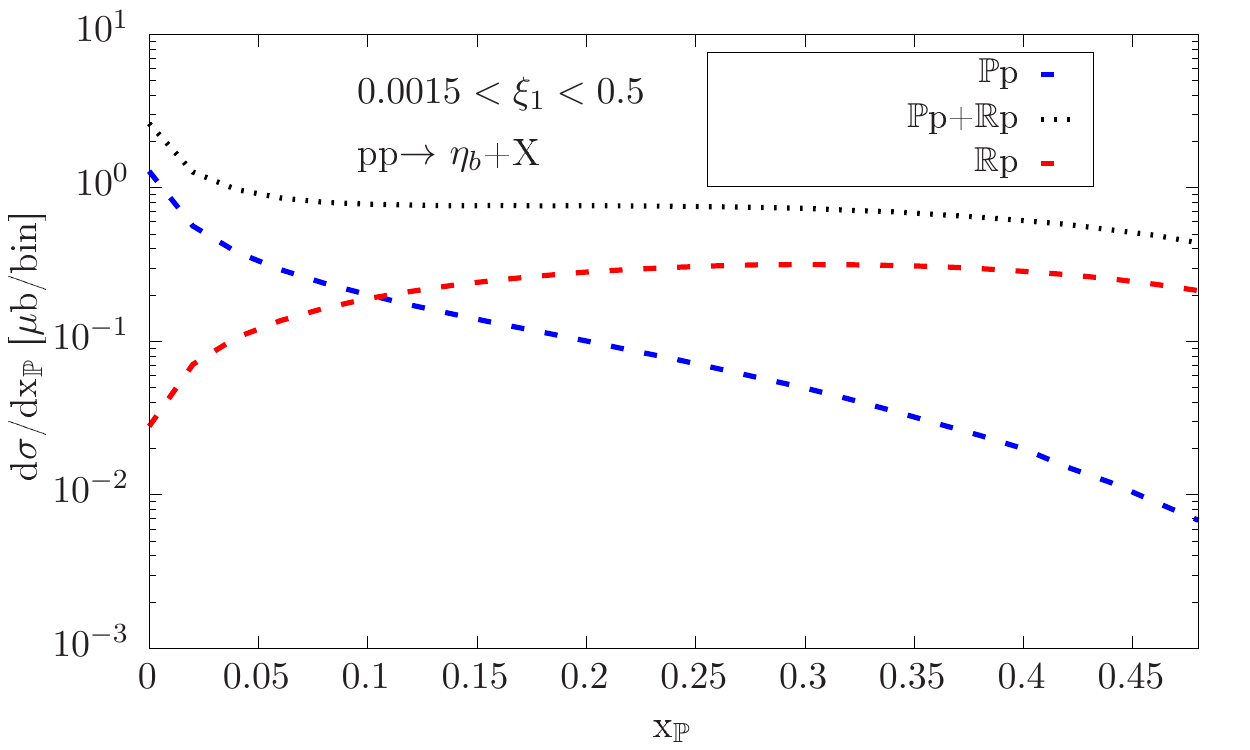}
%\end{minipage}
\caption{ \normalsize (color online)
The $x^{\eta_{c}}_{\mathbb{P}}$ and $x^{\eta_{b}}_{\mathbb{P}}$ distributions for the $\mathbb{P}$p (blue dashed line), $\mathbb{P}$p + $\mathbb{R}$p (black dotted line) and $\mathbb{R}$p (red dash dotted line) in SD processes.}
\label{fig7:limits}
\end{figure}

The presentation the $\beta$ distributions of $\eta_{c}$ and $\eta_{b}$ of three different cuts for forward detector acceptance is given in Fig.\ref{fig8:limits} for SD dissociation. The decreasing and fattening of the all  contributions are observed on plots. The $\mathbb{R}$p contribution shows a decreasing behavior along with $\beta$ and  is dominant for small $\beta$ over the $\mathbb{P}$p contribution. Nonetheless, we also notice that the $\mathbb{P}$p contribution overpasses that of $\mathbb{R}$p one for large $\beta$ and they become comparable for very small $\beta$. For the $0.015<\xi_{3}<0.15$ ($0.0015<\xi_{1}<0.5$), the $\mathbb{R}$p contribution is small compared to that of $\mathbb{P}$p for $\eta_{c}$($\eta_{b}$) for large $\beta$ and they become comparable for very small $\beta$.
\begin{figure}[htp]
\centering
%\begin{minipage}[t]{4.0cm}
\includegraphics[height=4.8cm,width=4.4cm]{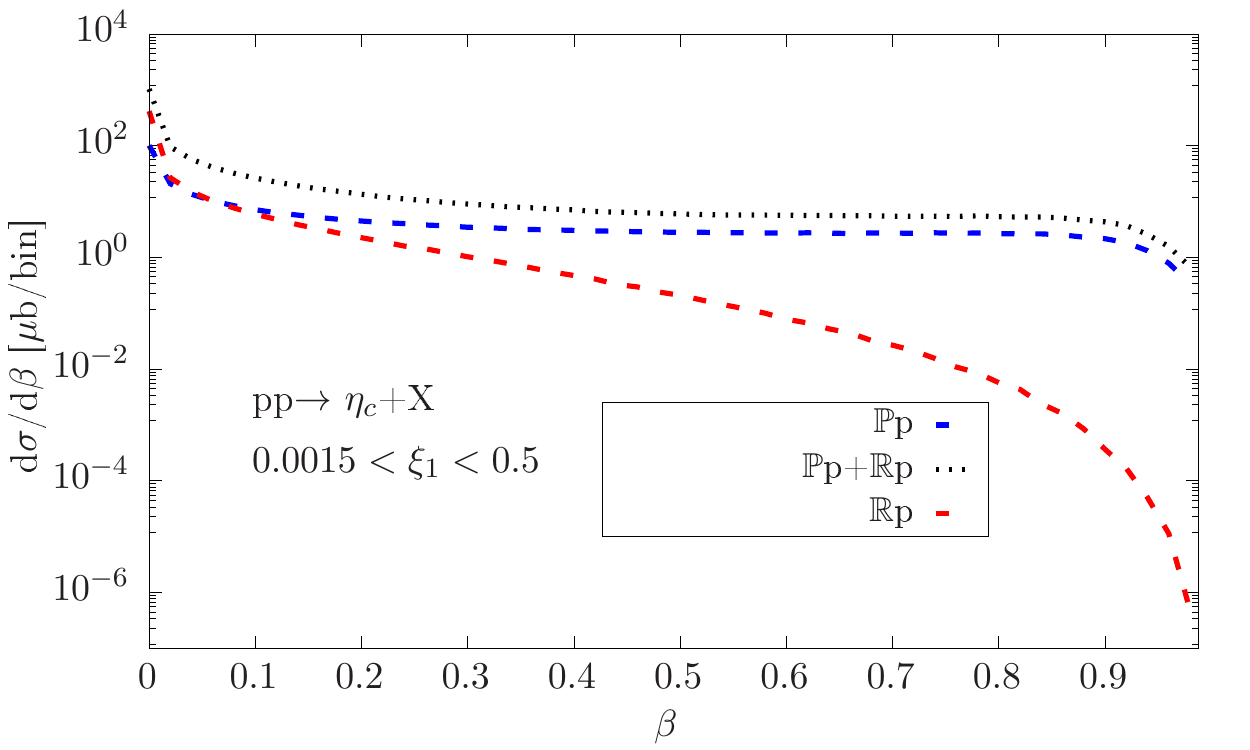}
\includegraphics[height=4.8cm,width=4.4cm]{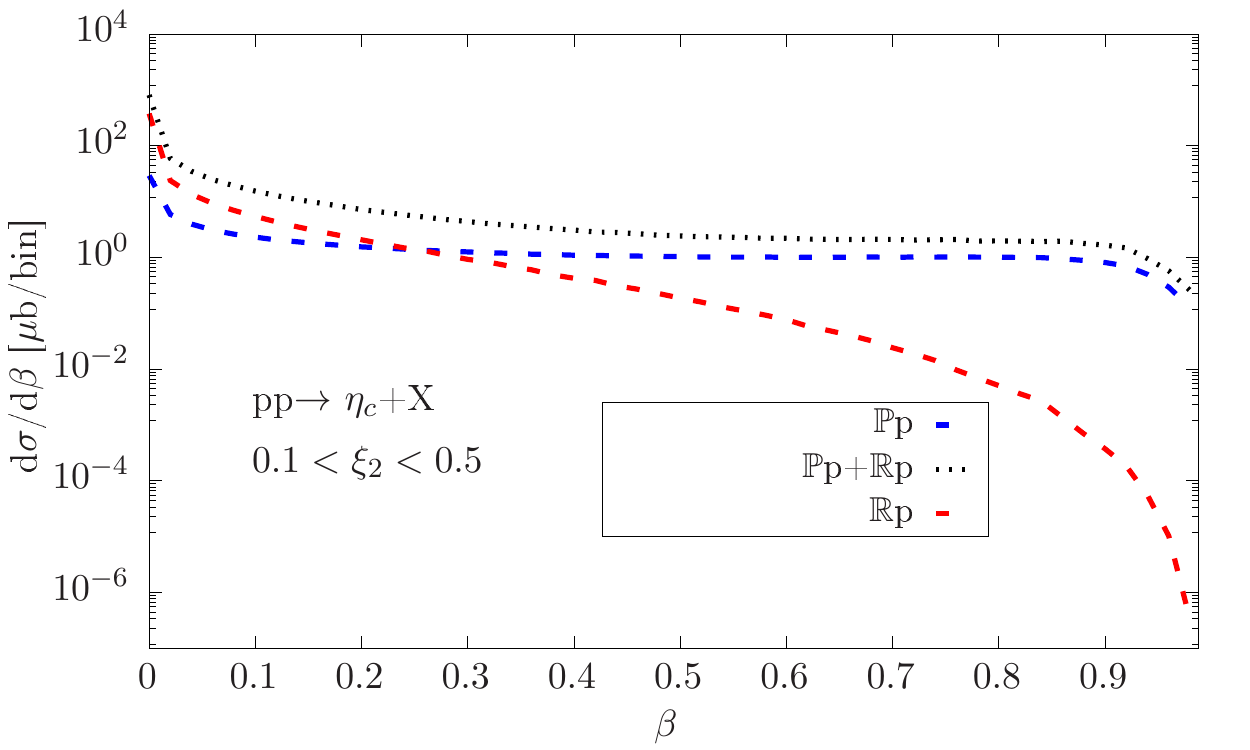}
\includegraphics[height=4.8cm,width=4.4cm]{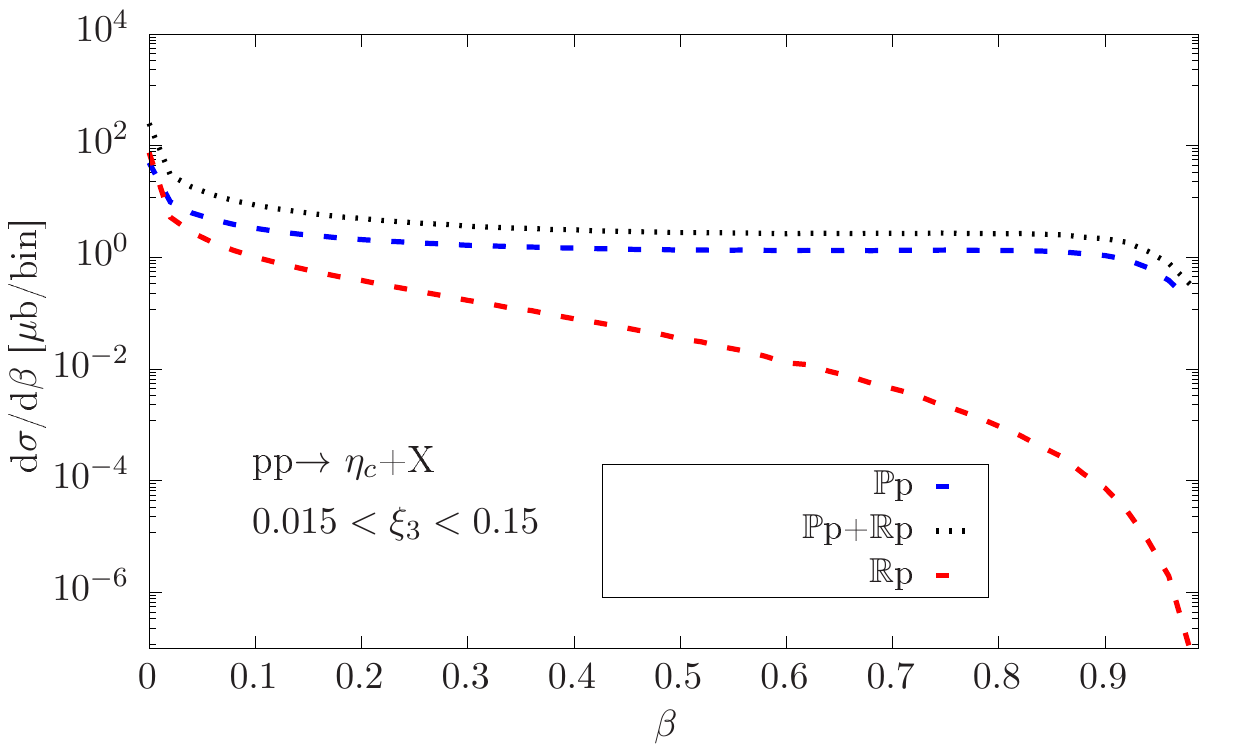}
\includegraphics[height=4.8cm,width=4.4cm]{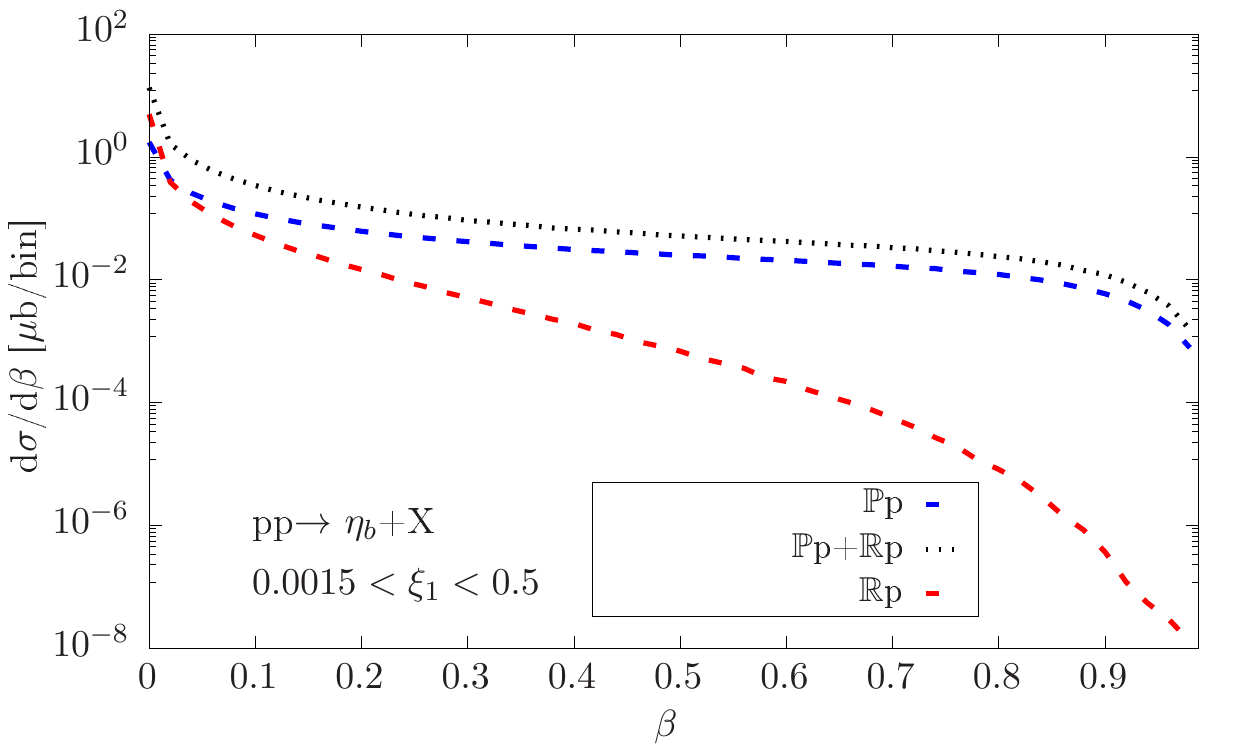}
%\end{minipage}
\caption{ \normalsize (color online)
The $\beta$ distributions  for the
$\mathbb{P}$p (blue dashed line), $\mathbb{P}$p+$\mathbb{R}$p(black dotted line) and  $\mathbb{R}$p (red dash dotted line) in SD processes .}
\label{fig8:limits}
\end{figure}

Our results show that $\eta_{c}$ and $\eta_{b}$ hadroproduction in SD and DD processes  at the LHC could be used to study the Reggeon contribution, since a kinematic window of dominance has been identified which could be used experimentally to isolate and constrain it. Our values are in agreement with the prediction that single diffractive cross-sections which should be approximately 10 times greater than in the double diffractive case \cite{Aad:2012pw,Marquet:2016ulz}. We have found that Reggeon exchanges contribute much more in some range of forward detector acceptance, and can almost never be completely disregarded. For large values of $x_{\mathbb{P}}$ and small values of $\beta$ but still within the detector acceptances, processes involving Reggeons can even dominate over the double-Pomeron exchange. For very small $\beta$, Reggeon and Pomeron exchanges are comparable.

\section{Summary and Conclusion }
\label{Conclusion}

In this work we calculate the hadroproduction of $\eta_{c}$ and日 $\rm \eta_{b}$ via single diffractive, double diffractive and non diffractive processes at the LHC $\rm \sqrt{s} = 13$ TeV energies. Considering the NRQCD formalism along with the resolved-Pomeron model, we predict the total cross sections, the differential and the production rates for these processes. Our results demonstrate that the contribution of Reggeon are non negligible orders of magnitude for certain forward detector acceptances, and therefore this study can be useful to better constrain the Reggeon parton content and correct the experimental model.

\begin{acknowledgments}
Hao Sun is supported by the National Natural Science Foundation of China (Grant No.11675033) and by the Fundamental Research Funds for the Central Universities (Grant No. DUT18LK27).
\end{acknowledgments}
\bibliography{v3}

\end{document}